\documentclass[fleqn,usenatbib]{mnras}

\usepackage{newtxtext,newtxmath}
\usepackage{amsmath}

\usepackage[T1]{fontenc}
\usepackage{ae,aecompl}

\usepackage{graphicx}	
\usepackage{amsmath}	

\usepackage{xcolor}
\usepackage{footnote}

\newcommand{\refeq}[1]{Eq.~(\ref{eq:#1})}          
          
\newcommand{\reffig}[1]{Fig.~\ref{fig:#1}}          
\newcommand{\refsec}[1]{Sec.~\ref{sec:#1}}
\newcommand{\refapp}[1]{App.~\ref{app:#1}}
\newcommand{\reftab}[1]{Tab.~\ref{tab:#1}}

\def\VEV#1{\left\langle #1 \right\rangle}
\def\rmd{\mathrm{d}}

\def\Msun{{\rm M}_\odot}

\begin{document}
\label{firstpage}
\pagerange{\pageref{firstpage}--\pageref{lastpage}}

\title[Microlensing of Star Clusters]{Statistical Microlensing Toward Magnified High-Redshift Star Clusters}



 
\author[L. Dai]{
Liang Dai$^{1,2}$\thanks{E-mail: liangdai@berkeley.edu}
\\
$^{1}$Institute for Advanced Study, 1 Einstein Drive, Princeton, NJ 08540, USA\\
$^{2}$University of California, Berkeley, Department of Physics, 366 LeConte Hall, Berkeley, CA 94720, USA.
}

\date{Accepted XXX. Received YYY; in original form ZZZ}


\maketitle

\begin{abstract}

We study light variability of gravitationally magnified high-redshift star clusters induced by a foreground population of microlenses. This arises as the incoherent superposition of light variations from many source stars traversing the random magnification pattern on the source plane. The light curve resembles a scale-invariant, Gaussian process on timescales of years to decades, while exhibits rapid and frequent micro-caustic crossing flares of larger amplitudes on timescales of days to months. For a concrete example, we study a young Lyman-continuum-leaking star cluster in the Sunburst Arc at $z=2.37$. We show that one magnified image happens to be intervened by a foreground galaxy, and hence should exhibit a variable flux at the $1$--$2\%$ level, which is measurable in space with $\sim 1$--$3\,$ks exposures on the Hubble Space Telescope and more easily with the James Webb Space Telescope, or even from the ground using a $\sim$4-meter telescope without adaptive optics. Detailed measurement of this variability can help determine the absolute macro magnification and hence the intrinsic mass and length scales of the star cluster, test synthetic stellar population models, and probe multiplicity of massive stars. Furthermore, monitoring the other lensed images of the star cluster, which are free from significant intervention by foreground microlenses, can allow us to probe planetary to stellar mass compact objects constituting as little as a few percent of the dark matter. Given the typical surface density of intracluster stars, we expect this phenomenon to be relevant for other extragalactic star clusters lensed by galaxy clusters.

\end{abstract}




\section{Introduction}
\label{sec:intro}

Strong gravitational lensing offers extraordinary opportunities to probe star formation at high redshifts. Thanks to lensing magnification, distant and intrinsically small star-forming galaxies can be detectable behind galaxy or galaxy cluster lenses, and often in the resultant giant arcs candidates of massive young star clusters are seen (e.g. ~\cite{Vanzella2017SuperStarCluster, rivera2019gravitational}). These impressive objects have recently aroused tremendous interest because they likely represent a major mode of star formation at Cosmic Noon in environments of low metallicity and high specific star-formation rate, and because their kind might be an efficient source of escaping ionizing radiation during the epoch of reionization~\citep{Ricotti2002ClusterReionization}. It is therefore worth considering how by exploiting lensing more information about their astrophysical properties might be extracted beyond the resolution limit of current instruments. 

When a population of compact objects intervene the line of sight, they act as microlenses and cast a random pattern of flux magnification on the source plane~\citep{GottIII1981QSOMicrolensing, Young1981QSOMicrolensing, chang1984star, Wambsganss2000CosmoMicrolensing}. Under these circumstances, individual members of a stellar system independently flicker as they steadily traverse this pattern, and they collectively give rise to a variable flux integrated over the entire association which is often unresolved or marginally resolved. This concept of statistical microlensing was previously proposed for ``pixel'' microlensing toward crowded star fields in nearby galaxies~\citep{Crotts1992M31, CrottsTomaney1996M31, AlcockEtAl1999DiffImage}, and for surface brightness variability of distant galaxies~\citep{LewisIbata2000MACHO, LewisIbataWyithe2000MACHO, Tuntsov2004ClusterMicrolensing, GilMerinoLewis2006StarFormingRegion}. 

In this work, we consider the similar effect acting on highly magnified star clusters behind galaxy cluster lenses. We turn our attention to the existing populations of microlenses -- stars either residing in intervening minor foreground galaxies or from the diffuse intracluster light (ICL)~\citep{zwicky1951coma, lin2004k, zibetti2005intergalactic}. These populations normally have a small mean convergence $\kappa_\star\sim 0.01$, and hence, toward unlensed or moderately magnified ($|\mu_{\rm macro}| \sim $ a few ) sources, would only cause insignificant microlensing under a low optical depth. Interestingly, for sources under high magnifications $|\mu_{\rm macro}| \sim \mathcal{O}(10)$--$\mathcal{O}(100)$, the strong external shear elevates microlensing into effectively the optical thick regime, with $\kappa_\star\,|\mu_{\rm macro}|=\mathcal{O}(1)$~\citep{2017ApJ...850...49V, 2018ApJ...857...25D, Oguri:2017ock}. Given the latest observational progress, it is timely to revisit statistical microlensing in this latter regime.

Recent years have seen important development in detecting individual superluminous stars in caustic-straddling arcs at $z\sim 1$--$1.5$~\citep{2018NatAs...2..334K, Chen:2019ncy, 2019ApJ...880...58K}, which are magnified by astounding factors in the hundreds and exhibited flux stochasticity due to microlensing. Arguably, observing statistical microlensing of a large association of bright stars can be complementary to, and perhaps in some aspects more advantageous, than observing similar effects on individually resolved stars. Foremost, highly magnified individual stars~\citep{1991ApJ...379...94M} become exceedingly difficult to detect beyond $z \simeq 2$ due to the Eddington limit on the intrinsic luminosity of massive stars~\citep{2018ApJ...867...24D}, while young star clusters at Cosmic Noon redshifts with lensing-enhanced visual magnitudes $m\simeq 22$--$25$ are much more easily detectable at high signal-to-noise ratios (SNRs). 

Individual stars are most easily detectable under the large magnification boost during micro-caustic crossings~\citep{Diego2019HighMuUniverse}. However, the erratic nature of these rapid events render them rare to be caught at random observing epochs. Moreover, for any single star, micro-caustic crossings often occur rather infrequently, unless in the improbable case of an exceptionally large $|\mu_{\rm macro}| \gtrsim 10^3$. Owing to the highly non-Gaussian nature of the light curve, an impractically long observing program would be necessary to accumulate a statistically meaningful amount of variability data. By contrast, these shortcomings can be significantly mitigated, if not completely overcome, by targeting a collective source consisting of a large number of stars. As we will show with examples, the collective variability resembles a scale-invariant, Gaussian process on the timescale of years, while on short timescales exhibits extremely frequent (e.g. a dozen of events per month) flares that arise from micro-caustic crossings of a large number of stars. The trade-off would be that a higher relative precision in photometry has to be achieved, which nevertheless might not be a problem --- after all, the absolute flux fluctuation from an incoherent sum of, say $\sim N$ comparably bright stars, is $\sim \sqrt{N}$ times larger than that of any single star. We therefore propose, in conjunction with the search and monitoring of extremely magnified individual stars (e.g. ~\cite{Kelly2019hstflashlight}), that microlensing-induced variabilities of magnified extragalactic star clusters be sought and measured.

With the goal to develop general insight into this phenomenon, in this work we examine the specific example of a Cosmic Noon star cluster found in the gravitationally lensed Sunburst Arc at $z=2.37$ behind a galaxy cluster lens~\citep{dahle2016discovery}, which has been dubbed the ``LyC Knot'' because Lyman Continuum emission was detected from it~\citep{rivera2019gravitational}. This LyC Knot shows a total of 12 lensed macro images, many of which likely to have high magnifications in the range of tens~\citep{vanzella2020ionizing}. Identified as a $3~$Myr-old compact star cluster with a total mass $M_\star \simeq 10^7\,(50/|\mu_{\rm macro}|)^{1/2}\,\Msun$~\citep{2019ApJ...882..182C}, microlensing induced flux variability seen across a wide range of observed wavelengths must be dominated by $\sim 10^3$--$10^4$ O-type main-sequence stars and evolved B-type supergiants. Assuming the standard initial mass function (IMF) and neglecting stellar multiplicity, our stellar population modeling suggests that one of the lensed images of the LyC Knot, due to an intervening minor foreground galaxy G1 (which provides $\kappa_\star \simeq 0.01$ as we estimate), should exhibit a variable flux at $ 0.8$--$2.5\%$ level if it has a macro magnification $|\mu_{\rm macro}|=20$--$50$. Given the remarkable apparent brightness of those lensed images, such percent-level variability should be measurable in the rest-frame UV bands for $\sim 1$--$3\,$ks exposures with the HST~\citep{rivera2019gravitational}.

We will show that, as long as the magnified flux is held fixed by observation, the typical amplitude of flux variability depends only weakly on the IMF and metallicity, but is sensitive to the macro magnification $\mu_{\rm macro}$. If accurate modeling of the microlens surface density $\kappa_\star$ is available, measuring the flux variability of a star cluster can determine its absolute mass scale by determining $|\mu_{\rm macro}|$, which will be a valuable result since lens modeling may suffer from large uncertainty at the locations of high-magnification features~\citep{Priewe2017LensModelMuError}. Detailed statistical quantification of the flux variability under statistical microlensing and comparing data to theory will offer a new avenue to test stellar population synthesis in the context of high-$z$ environments of high specific star-formation rate and low metallicity. We furthermore suggest that the fraction of massive stars in binary or multiple stellar systems~\citep{Sana2012MassiveBinaries, Sana2013MultipleOStars}, another crucial aspect of population synthesis particularly important for understanding ionizing fluxes~\citep{Stanway2016BinaryReionization, Ma2016BinaryReionization, Rosdahl2018binary}, may also be probed with statistical microlensing, as a larger fraction will enhance the variability amplitude.

Many of the earlier works on statistical microlensing were originally motivated by the prospect to probe stellar or planetary mass compact objects which were thought to constitute the cosmological Dark Matter (DM). Since then, an $\mathcal{O}(1)$ mass fraction of these compact DM has been ruled out by microlensing surveys~\citep{alcock2001macho, griest2013new, tisserand2007limits, 2019NatAs...3..524N} and various other astrophysical~\citep{Brandt2016MACHOconstraint} and cosmological tests~\citep{mediavilla2017limits, AliHaimoud2017machoCMB, zumalacarregui2018limits}. In spite of that, we propose that empirical constraints on percent-level variability seen in magnified star clusters (e.g. with the other lensed images of the LyC Knot in the Sunburst Arc) could tighten the limit on the mass fraction to $\mathcal{O}(10^{-2})$ across the mass range $\sim 10^{-6}$--$10^2\,\Msun$.

The remainder of this paper is organized as follows. In \refsec{theory}, we justify several simplifying assumptions underlying our statistical description of the collective microlensing phenomenon, and introduce the formalism to quantify the amplitude variance and temporal correlation of flux variability. Then in \refsec{sunburst}, we study the specific case of the LyC Knot of the Sunburst Arc. In \refsec{disc}, we discuss related issues regarding variable stars, stellar multiplicity, and compact dark matter. Finally, we summarize our conclusions in \refsec{concl}. Additional technical details are presented in Appendices: we discuss in \refapp{clusterhalo} a simple mass profile model for the cluster lens in front of the Sunburst Arc, and estimate the abundance of intracluster stars; we characterize in \refapp{G1} the stellar population of the foreground galaxy G1, from which we derive the abundance of microlenses toward one magnified image of the LyC Knot.

\section{Theoretical Formalism}
\label{sec:theory}

The overall flux of a star cluster can have temporal variability if individual member stars have variable fluxes due to microlensing~\citep{2017ApJ...850...49V, 2018ApJ...857...25D, Oguri:2017ock}
When the star cluster is unresolved, what is measurable is the integrated flux but not the fluxes from individual members~\citep{Crotts1992M31, LewisIbataWyithe2000MACHO, LewisIbata2000MACHO}. Recently, collective variability was explored in \cite{Dai2020S1226} under the idealistic assumption of identical member stars, in the context of understanding highly magnified, asymmetric image pairs of compact star clusters residing in the caustic straddling lensed galaxy SGAS J122651.3+215220~\citep{dahle2016discovery}. 

In this work, we expand that study by accounting for a realistic source flux distribution based on synthetic stellar populations. A general theoretical framework to quantify statistical microlensing was presented in \cite{Tuntsov2004ClusterMicrolensing}. In the following, we follow closely the reasoning of that reference and reproduce many of results there, albeit we will employ the method of characteristic function to understand Gaussianization of light variability.

We consider, in the projected vicinity of the line of sight toward the star cluster, randomly located microlenses, which are superimposed on top of some locally uniform coarse-grained convergence $\kappa_0$ and shear $\gamma_0$ from the smooth macro lens. The microlenses on average make a small but important contribution $\kappa_\star=\Sigma_\star/\Sigma_{\rm crit}\sim 10^{-2}$ to $\kappa_0$, where $\Sigma_\star$ is their surface mass density, and $\Sigma_{\rm crit}=(c^2/4\,\pi\,G)\,(D_S/D_L\,D_{LS})$ is the critical surface density~\citep{blandford1986fermat}, with $D_L$, $D_S$ and $D_{LS}$ being the angular diameter distances to the lens plane, to the source plane, and from the lens plane to the source plane, respectively. The star cluster is subject to a (signed) macro magnification $\mu_{\rm macro} = \mu_t\,\mu_r$, where \begin{align}
\label{eq:mutmur}
    \mu_t=(1-\kappa_0 - \gamma_0)^{-1},\quad {\rm and} \quad \mu_r=(1-\kappa_0 + \gamma_0)^{-1}
\end{align}
are the one-dimensional magnifications along and perpendicular to the macro elongation, respectively. We will focus on the regime of large macro magnification near a tangential caustic, i.e. $|\mu_t|\gg 1$ and $\mu_r \sim \mathcal{O}(1)$.

\subsection{Basic Assumptions}

To begin with, we justify several simplifying assumptions we will make in this paper. 

Firstly, we assume that individual member stars have uncorrelated variabilities. For stellar microlenses of masses $\sim \mathcal{O}(1\,\Msun)$, this should be a good approximation because the characteristic length scale of the magnification pattern cast onto the source plane, on the order of the Einstein scale $\sim \mathcal{O}(10^3\,{\rm AU})$ (or more precisely a factor of $\sqrt{|\mu_{\rm t}|}$ larger perpendicular to macro elongation~\citep{Oguri:2017ock}), is smaller than the typical projected separations between the brightest member stars, say $\sim 10^4$--$10^5$ OB stars packed within a radius of a few ${\rm pc}$. This assumption becomes invalid for tight binary or multiple stars, which are in fact not uncommon for massive stars~\citep{sana2010multiplicity, sana2016multiplicity}. For now, we neglect stellar multiplicity, and will comment on this later.

Secondly, the source-plane magnification pattern is assumed to be statistically homogeneous. For microlenses in either the intracluster space or from an old stellar population in a foreground galaxy, their projected positions typically have randomized sufficiently. Also, the extent of the entire star cluster (a few pc across), when mapped onto image plane, probes typically $\sim 10^3$--$10^4$ microlenses simultaneously, adequately sampling the underlying statistics.

For a third premise, we assume that all member stars traverse the magnification pattern at the same transverse velocity, which we define to be at an angle $\phi$ with respect to the direction of macro elongation. This uniform velocity is essentially set by the transverse motion of the star cluster as a whole relative to the lens. After all, internal motions of the star cluster members $\sim \mathcal{O}(10\,{\rm km/s})$ are much slower than the typical peculiar motion of the large scale structure $\sim \mathcal{O}(10^2$--$10^3\,{\rm km/s})$~\citep{1991ApJ...379...94M, 2017ApJ...850...49V}. The velocity dispersion of the microlenses are also negligible, as it is suppressed by a factor of $|\mu_t|$ when mapped onto the source plane.

Since in this context source stars are effectively point sources for magnifications $\lesssim 10^4$~\citep{2017ApJ...850...49V}, we assume that all member stars, across a wide range of wavelengths, have the same statistical distributions for the fractional flux variability.

For one more simplifying assumption, we will neglect any intrinsic variability of the source stars including stellar pulsations, outbursts, or other explosive transients (see \refsec{intrinsic_variable_star} for discussion). We will also neglect the effects of eclipses in multiple-star systems. With these assumptions, in the following we will analytically express the statistics of the collective flux variability in terms of that for individual member stars, following the standard results in statistics.

\subsection{Individual Stars}

Under microlensing, the flux $F$ of a given individual star fluctuates around the mean value $\bar{F}$,
\begin{align}
    F = \bar{F}\,\left( 1 + \delta \right).
\end{align}
Consider the probability distribution function (PDF) for the fractional fluctuation $\delta$, $P(\delta)$, which satisfies $\int\,\rmd\delta\,P(\delta) = 1$ and $\langle\delta\rangle = \int\,\rmd\delta\,P(\delta)\,\delta = 0$. The characteristic function (CF) is given by the Fourier transform of $P(\delta)$,
\begin{align}
\label{eq:cf}
    \Phi(\omega) \equiv \langle e^{i\,\omega\,\delta} \rangle = \int\,\rmd \delta\,P(\delta)\,e^{i\,\omega\,\delta}.
\end{align}
The logarithm of the CF, $\Psi(\omega)=\ln\Phi(\omega)$, gives the cumulant generating function (CGF).

For an idealized point source, $P(\delta)$ peaks at $\delta \sim \mathcal{O}(1)$, but has a power-law tail $P(\delta) \propto \delta^{-3}$ for $\delta \gg 1$~\citep{peacock1982gravitational, 1987ApJ...319....9S} because fold singularities dominate the PDF in the high magnification regime~\citep{blandford1986fermat, SchneiderWeiss1988LightProp}. Formally, this results in a logarithmic divergence in the second-order cumulant $\langle\delta^2\rangle_c=\langle \delta^2 \rangle - \langle \delta \rangle^2$, and power-law divergences in higher order cumulants. For a realistic stellar source, the finite source size truncates the high magnification tail and hence regularizes these divergences, but the cumulants may still have large numerical values, reflecting the strongly non-Gaussian nature of $P(\delta)$. When all cumulants $\langle \delta^n \rangle_c,\,n=2,3,\cdots$ are finite, the CGF is analytic around $\omega=0$, and the values of the cumulants can be read from the Taylor expansion coefficients:
\begin{align}
\label{eq:cgf}
    \Psi(\omega) = \sum^\infty_{n=2}\,\frac{i^n}{n!}\,\langle \delta^n \rangle_c\,\omega^n.
\end{align}
In particular, $\langle \delta^2 \rangle_c$ is equivalent to the quantity $\varepsilon^2_\mu$ defined by \cite{Tuntsov2004ClusterMicrolensing}.

\subsection{Integrated Flux of a Cluster: One-point statistics}

We now consider the many members of a star cluster. The members can be partitioned into bins according to the mean flux $\bar{F}$, labeled by $I=1,2,\cdots$. The $I$-th bin consists of $N_I$ (nearly) identical stars, each of which has a mean flux $\bar{F}_I$. The flux integrated over the entire cluster therefore has a mean value $\bar{F}_{\rm cl}=\sum_I\,N_I\,\bar{F}_I$. The integrated flux $F_{\rm cl}$ at a given epoch fluctuates around this value by a fractional amount $\Delta$, defined through
\begin{align}
    F_{\rm cl} = \bar{F}_{\rm cl}\,\left( 1 + \Delta \right).
\end{align}
Under our assumptions, $\Delta$ has a vanishing mean $\langle \Delta \rangle = 0$, and its own CGF is given by
\begin{align}
    \Psi_{\rm cl}(\omega) = \ln\Phi_{\rm cl}(\omega) = \ln\langle e^{i\,\omega\,\Delta} \rangle =  \sum_I\,N_I\,\Psi\left( \frac{\bar{F}_I}{\bar{F}_{\rm cl}}\,\omega \right).
\end{align}
Taking the limit of a continuous distribution of $\bar{F}$ for individual stars, we introduce the flux distribution function $\rmd N/\rmd \bar{F}$, and derive
\begin{align}
    \bar{F}_{\rm cl} = \int\,\rmd \bar{F}\,\frac{\rmd N}{\rmd \bar{F}}\,\bar{F},
\end{align}
and
\begin{align}
\label{eq:Psicl}
    \Psi_{\rm cl}(\omega) = \int\,\rmd\bar{F}\,\frac{\rmd N}{\rmd \bar{F}}\,\Psi\left( \frac{\bar{F}}{\bar{F}_{\rm cl}}\,\omega \right)
\end{align}
Inserting \refeq{cgf} into \refeq{Psicl}, we have
\begin{align}
    \Psi_{\rm cl}(\omega) = \sum^{\infty}_{n=1}\,\frac{i^n}{n!}\,\left( \frac{1}{\bar{F}_{\rm cl}^n} \int\,\rmd\bar{F}\,\frac{\rmd N}{\rmd \bar{F}}\,\bar{F}^n \right)\,\langle \delta^n \rangle_c\,\omega^n.
\end{align}
Therefore, $\Delta$ has cumulants
\begin{align}
\label{eq:Deltacumu}
    \langle \Delta^n \rangle_c = \frac{1}{N^{n-1}} \frac{\langle \bar{F}^n \rangle}{\langle \bar{F} \rangle^n}\,\langle \delta^n \rangle_c,
\end{align}
where we define the population averaged moments of the mean flux $\bar{F}$,
\begin{align}
    \langle \bar{F}^n \rangle \equiv \frac{1}{N}\, \left( \int\,\rmd\bar{F}\,\frac{\rmd N}{\rmd \bar{F}}\,\bar{F}^n \right),
\end{align}
and $N\equiv\int\,\rmd \bar{F}\,(\rmd N/\rmd \bar{F})$ is the total number of stars. From \refeq{Deltacumu}, we find
\begin{align}
    \frac{\left[ \langle \Delta^n \rangle_c \right]^{1/n}}{\left[ \langle \Delta^2 \rangle_c \right]^{1/2}} = N^{\frac{1}{n}-\frac12}\,\frac{\langle \bar{F}^n \rangle^{1/n}}{\langle \bar{F}^2 \rangle^{1/2}}.
\end{align}
If the distribution $\rmd N/\rmd \bar{F}$ is fixed but $N$ increases, the higher order cumulants $\langle \Delta^n \rangle_c$ for $n=3,\,4,\,\cdots$ are all suppressed relative to the second order cumulant $\langle \Delta^2 \rangle_c$. In the limit that all cumulants $\langle \Delta^n \rangle_c$ become negligibly small except for $n=2$, $P(\Delta)$ approaches a normal distribution with a variance
\begin{align}
\label{eq:rmsflux}
    \langle \Delta^2 \rangle_c = \epsilon^2_2\, \langle \delta^2 \rangle_c,
\end{align}
where we introduce the dimensionless parameter independent of microlensing:
\begin{align}
\label{eq:eps2}
    \epsilon_2 \equiv \left( \frac{1}{N}\,\frac{\langle \bar{F}^2 \rangle}{\langle \bar{F} \rangle^2}. \right)^{1/2}.
\end{align}
\refeq{rmsflux} states that the fractional fluctuation in the cluster integrated flux $F_{\rm cl}$ has a root mean square (RMS) that is a factor $\epsilon_2$ smaller than that of the individual stars. For a stellar population of fixed intensive properties, $\epsilon_2$ scales with the total stellar mass $M_\star$ as $\epsilon_2 \propto M^{-1/2}_\star$. This is the intuitive result that the highly non-Gaussian flux variability of a large number of cluster member stars collectively give rise to variability in the integrated flux that is more Gaussianized and has a reduced amplitude.

At low microlensing optical depths $\kappa_\star\,|\mu_{\rm macro}| \ll 1$, the high magnification tail of $P(\delta)$ only makes a subdominant contribution to $\langle \delta^2 \rangle_c$, so the RMS of $\Delta$ is primarily determined by the width of the core around $\delta \approx 0$.

At high microlensing optical depths $\kappa_\star\,|\mu_{\rm macro}| \gtrsim 1$, a situation we will demonstrate later, it is the heavy tail of the PDF $P(\delta)$ at $\delta \gg 1$ rather than the core at $\delta \approx 0$ that dominates $\langle \delta^2 \rangle_c$. In this situation, the RMS of $\Delta$ primarily depends on micro-caustic crossings of individual stars. By the same logic, any residual non-Gaussian behavior in $\Delta$ is sensitive to the high-$\delta$ cutoff in $P(\delta)$. 

\subsection{Integrated Flux of a Cluster: Two-point Statistics}

The one-point PDF $P(\delta)$ or $P(\Delta)$ does not capture all statistical information. Additional information is encoded in multi-point statistics of $\delta$, namely correlation between $\delta$'s measured at multiple source-plane positions. These are directly translated into correlation between $\delta$'s measured at different epochs under a uniform source motion, and hence provide the timescale information of the variability~\citep{WyitheTurner2002quasarGRBvariability}. Previously, \cite{GeraintIrwin1995TemporalAnalysis} and \cite{Neindorf2003MicrolensingAutoCorr} studied the temporal correlation in the light curve of microlensed quasars. Following the same logic, we can study the temporal correlation of the integrated flux $F_{\rm cl}$. 

The simplest example is the two-point correlation function,
\begin{align}
    \xi(\vartheta) \equiv \langle \delta(\theta)\,\delta(\theta + \vartheta) \rangle, 
\end{align}
also referred to as the structure function by \cite{wyithe2001determining} (and a related definition by \cite{GeraintIrwin1995TemporalAnalysis}). By statistical homogeneity of the source-plane magnification pattern, it is only dependent on the source-plane separation $\vartheta$ along the source trajectory, or equivalently on the time separation under uniform source motion. Define the Fourier transform,
\begin{align}
    \tilde \delta(f) \equiv \int\,\rmd\theta\,e^{i 2\pi f \theta}\,\delta(\theta),
\end{align}
where $f$ is the spatial frequency conjugate to $\theta$. The Fourier transform has a power spectrum
\begin{align}
    \langle \tilde\delta(f)\,\tilde\delta^*(f') \rangle = P_\delta(f)\,\delta_D(f - f'),
\end{align}
where $P_\delta(f)=\int\,\rmd\vartheta\,e^{i 2\pi f \vartheta}\,\xi(\vartheta)$. 

In a similar way, we can define the power spectrum $P_\Delta(f)$ for the integrated flux variability. Under the assumptions of independent variability and uniform source motions among individual stars, we have the analogy of \refeq{rmsflux} for the power spectra
\begin{align}
\label{eq:PDelf}
    P_\Delta(f) = \epsilon^2_2\,P_\delta(f).
\end{align}
Without explicitly writing them down, we point out that the disconnected part of the higher-order multi-point statistics also satisfy various relations that resemble \refeq{Deltacumu}. 

According to \refeq{PDelf}, $P_\Delta(f)$ inherits exactly the same functional shape as $P_\delta(f)$, despite the fact that flux variabilities of individual stars are highly non-Gaussian while that of the integrated flux Gaussianizes substantially. This feature implies, as long as our basic assumptions remain valid, that by measuring the shape of $P_\Delta(f)$ (and other disconnected multi-point correlation statistics) one can extract statistical information about the variability of individual stars, which reflect the population properties of the microlenses.

\section{Case study: LyC Knot in Sunburst Arc}
\label{sec:sunburst}

We now apply our theoretical framework to an observed star cluster that we judge is subject to statistical microlensing. It is a compact young star cluster at $z=2.37$, whose host galaxy is gravitationally distorted into a giant arc, dubbed the Sunburst Arc~\citep{RiveraThorsen2017bSunburst}, by a foreground galaxy cluster PSZ1 G311.65-18.48 (hereafter PSZ1-G311 for short)~\citep{2014A&A...571A..29P}. The star cluster is found to show an impressive set of 12 magnified images along the arc. Remarkably, redshifted LyC radiation was directly detected from each of these avatars~\citep{rivera2019gravitational}. Referred to as the LyC Knot, this star cluster is intriguing in its own right as it may give clues about the astrophysical properties of ionizing sources at high redshifts.

At the projected radii along the Sunburst Arc, we set a fiducial value $\kappa_0=0.6$ based on a spherical Navarro-Frenk-White (NFW) model~\citep{1996ApJ...462..563N, navarro1997universal} for the cluster lens profile. See \refapp{clusterhalo} for details. Given the values of $\kappa_0$ and $\mu_{\rm macro}$, we separately determine $\mu_t$ and $\mu_r$ from \refeq{mutmur}.

Stellar population modeling indicates that, out to the rather large projected distance of the Sunburst Arc from the brightest cluster galaxy (BCG), $\sim 170\,$kpc, the ICL contributes a small convergence from stellar microlenses $\kappa_\star \lesssim 0.002$ (see \refapp{clusterhalo}), which as we can tell from later analysis is unimportant. Therefore, microlensing by intracluster stars is unlikely to play a significant role in the case of the Sunburst Arc, unlike other studied giant arcs with $\kappa_\star = \mathcal{O}(10^{-2})$ in different massive lensing galaxy clusters, e.g. MACS J1149.5+2223~\citep{Oguri:2017ock}, MACS J0416.1-2403~\citep{2019ApJ...880...58K}, and SDSS J1226+2152~\citep{Dai2020S1226}. 

Interestingly, a faint foreground galaxy, which we refer to as G1, happens to intervene the line of sight toward one of the 12 lensed images of the Lyc Knot --- Image 5 following the denotation of \cite{rivera2019gravitational} (see \reffig{arc}). The expected surface number density of stellar microlenses in G1, $\kappa_\star\sim \mathcal{O}(10^{-2})$, together with the (most likely) high macro magnification factor of Image 5, $|\mu_{\rm macro}| \sim \mathcal{O}(10)$, makes this a realistic case for statistical microlensing. Our goal will be to quantify the resultant flux variability, in terms of the variance as in \refeq{rmsflux}, and the power spectrum as in \refeq{PDelf}.

\subsection{Stellar Source Population of the LyC Knot}

\begin{figure*}
	\includegraphics[scale=0.39]{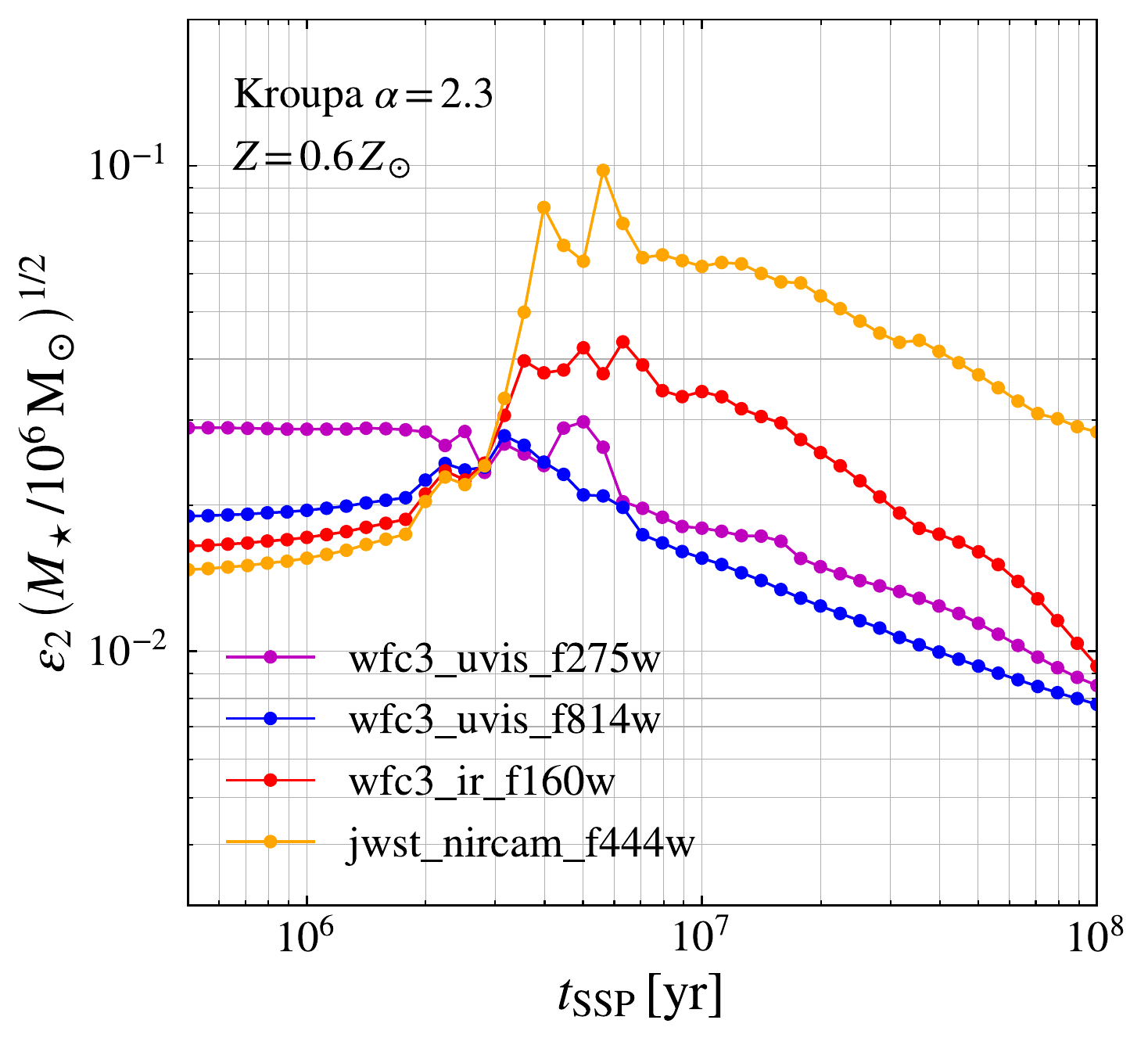}
	\includegraphics[scale=0.39]{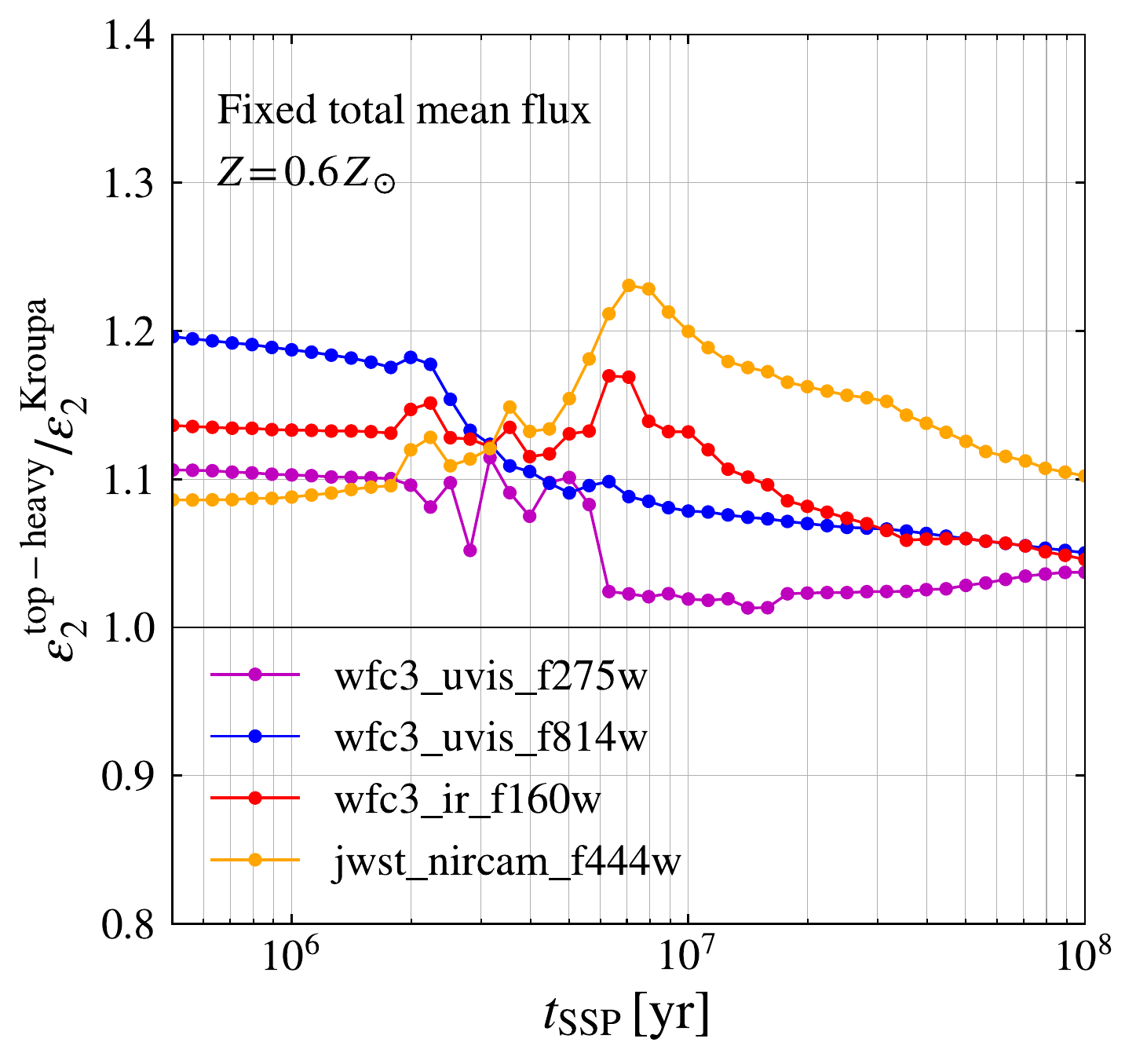}
	\includegraphics[scale=0.39]{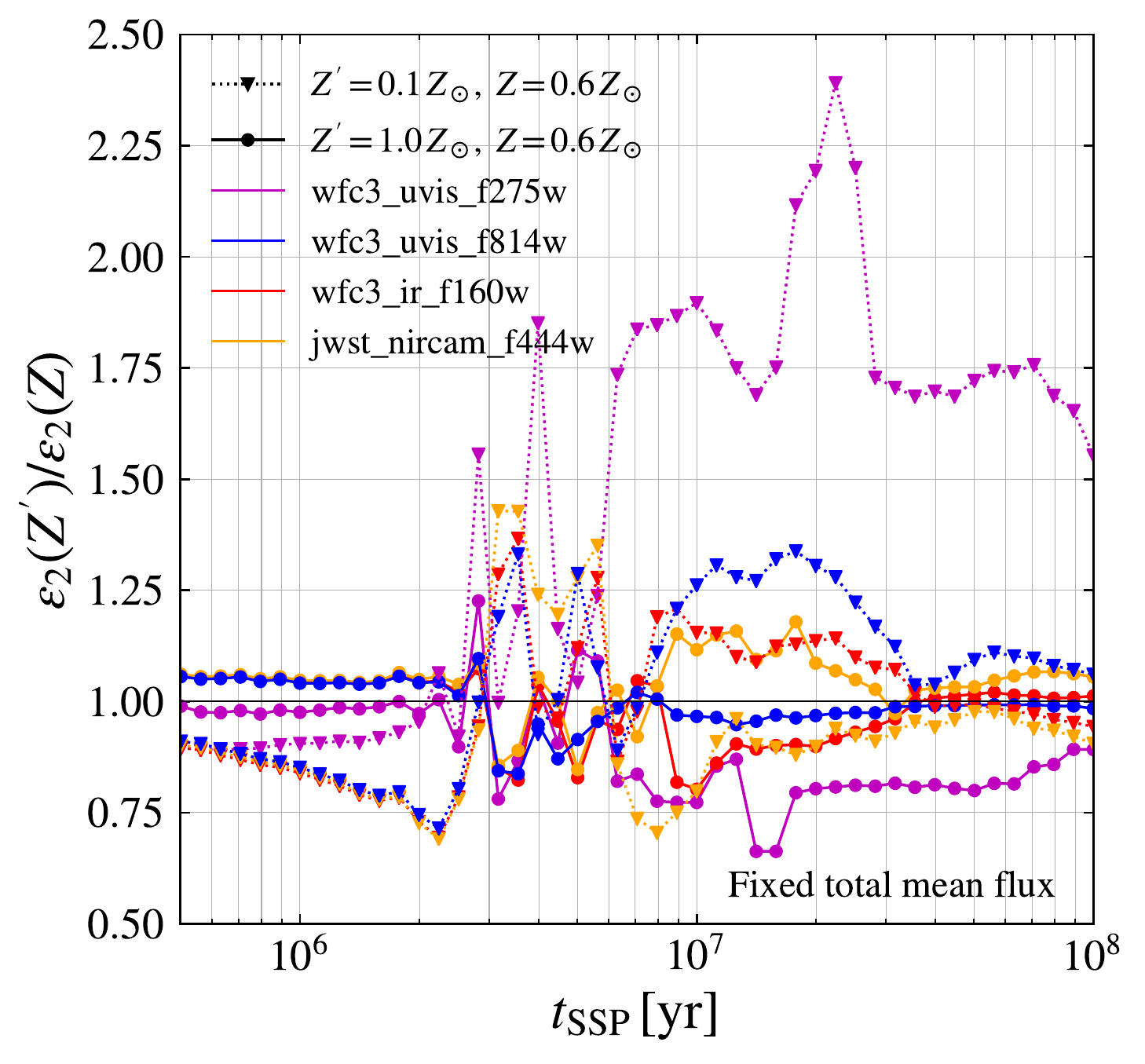}
    \caption{ Theoretical values for the parameter $\epsilon_2$ computed using the stellar population synthesis package FSPS~\citep{conroy2009propagation, conroy2010propagation}, accounting only for single stellar evolution. We consider the evolution of an SSP with age $t_{\rm SSP}$. {\bf Left panel}: The parameter $\epsilon_2$ normalized to a total stellar mass $M_\star=10^6\,\Msun$ for the double power-law Kroupa IMF ($\alpha=2.3$ for $M>0.5\,\Msun$) as a function of $t_{\rm SSP}$. {\bf Middle panel}: Ratio in $\epsilon_2$ between the Kroupa IMF and a top-heavy IMF ($\alpha=1.6$ for $M>0.5\,\Msun$) as a function of $t_{\rm SSP}$. At fixed integrated mean flux $\bar{F}_{\rm cl}$, $\epsilon_2$ only depends weakly on the IMF slope for massive stars. {\bf Right panel:} Ratio in $\epsilon_2$ between different metallicity choices as a function of $t_{\rm SSP}$. We compare $Z=0.1\,Z_\odot$ and $1\,Z_\odot$ to the fiducial metallicity $Z=0.6\,Z_\odot$ found for the LyC Knot~\citep{2019ApJ...882..182C}. At fixed integrated mean flux $\bar{F}_{\rm cl}$ and for stellar populations younger than $\sim 6\,$Myr, $\epsilon_2$ only depends mildly on metallicity. In all panels, we show curves for three HST wide filters, F275W, F814W and F160W, as well as one JWST NIRCAM wide filter F444W, respectively, all computed for the source redshift $z_s=2.37$ of the Sunburst Arc. These filters are chosen as examples to cover UV, optical, and IR wavelength ranges in the observer frame. We note that an age $t_{\rm SSP}=3.0$--$3.6\,$Myr was reported for the LyC Knot with or without including binary stellar evolution~\citep{2019ApJ...882..182C}.}
    \label{fig:eps2}
\end{figure*}

To evaluate one ingredient in \refeq{rmsflux}, the parameter $\epsilon_2$, we have to model the population of source stars within the LyC Knot. Detailed rest-UV spectroscopy carried out by \cite{2019ApJ...882..182C} indicates that the LyC Knot is a Cosmic Noon cousin of the young star cluster R136 situated at the center of the star forming region 30 Doradus in the Large Magellanic Cloud (LMC)~\citep{crowther2016r136}. It is however one or two orders of magnitude more massive with a total stellar mass $M_\star \sim 10^6$--$10^7\,\Msun$, while still has a compact spatial extent $\lesssim 20\,$--$30\,$pc~\citep{vanzella2020ionizing}. Specifically, \cite{2019ApJ...882..182C} reported an age $t_{\rm SSP} = 3.0$--$3.6\,$Myr, metallicity $Z = 0.55$--$0.66\,Z_\odot$, and inferred source-frame dust extinction $E(B-V)\simeq 0.15$. 

At such a young age, the dominant contributors to the rest-frame UV/optical fluxes are massive O-type main-sequence stars and evolved B-type supergiants. The lower number of these massive stars compared to that of the low-mass stars has the favorable implication that the parameter $\epsilon_2$ may not be tremendously suppressed to an observationally inaccessible level. 

To evaluate $\epsilon_2$, we construct synthetic simple stellar populations (SSPs) at a uniform age $t_{\rm SSP}$ using the public code Flexible Stellar Population Synthesis (FSPS)~\citep{conroy2009propagation, conroy2010propagation}. With FSPS, we adopt a double power-law model for the IMF including initial masses between $0.08\,\Msun$ and $100\,\Msun$. For initial masses between $0.08\,\Msun$ and $0.5\,\Msun$, we fix the slope to be $\alpha=1.3$. For initial masses heavier than $0.5\,\Msun$, we consider different scenarios: (1) a standard Salpeter-like slope $\alpha = 2.3$~\citep{kroupa2001variation}; (2) a top-heavy slope $\alpha = 1.6$, which has been suggested for intense star forming regions under extreme conditions~\citep{jevrabkova2017formation}. Besides, the parameter $\epsilon_2$ is also dependent on metallicity. We also note that binaries or multiple stellar systems are not included in these models.

In \reffig{eps2}, we show $\epsilon_2$ as a function of the stellar age $t_{\rm SSP}$. Since $\epsilon_2$ is filter dependent, we evaluate $\epsilon_2$ for several HST and JWST wide filters that span the observed wavelength range from UV to optical and near-IR, redshifted from $z_s = 2.37$. For the standard high-mass IMF slope $\alpha=2.3$, and $Z = 0.6\,Z_\odot$ inferred for the LyC Knot, we find $\epsilon_2 \approx ( 1.5\textrm{--}3 )\%\,( M_\star/10^6\,\Msun )^{-1/2}$ for stellar populations younger than $\sim 3\,$Myr and in a variety of (observer-frame) filters from UV to IR (left panel of \reffig{eps2}). The value of $\epsilon_2$ begins to evolve significantly when the star cluster age further. In the bluer filters F275W and F814W, $\epsilon_2$ slowly transitions to a monotonic decrease for $t_{\rm SSP}\gtrsim 6\,$Myr, while in the redder filters F160W and F444W, $\epsilon_2$ first reaches a peak of a factor of a few to ten larger at $t_{\rm SSP} \simeq 3$--$6\,$Myr, and then steadily decreases for $t_{\rm SSP} \gtrsim 10\,$Myr. 

Across a wide range of wavelengths, very young stellar populations with $t_{\rm SSP} \lesssim 3\,$Myr exhibit a stable value of $\epsilon_2$ dominated by massive O stars during their main sequence stage. As the stellar population grow older, many massive stars become evolved ones and are subject to dramatic changes in their stellar radii and temperatures. Consequently, the flux distribution $\rmd N/\rmd \bar{F}$ significantly changes and so does the value of $\epsilon_2$.

The middel panel of \reffig{eps2} shows that $\epsilon_2$ only mildly depends on the IMF slope for massive stars for a wide range of ages $t_{\rm SSP} < 100\,$Myr, as long as the integrated mean flux $\bar{F}_{\rm cl}$ of the star cluster is kept fixed (by observation). Between the standard scenario for the IMF slope and the more extreme top-heavy scenario, $\epsilon_2$ varies by less than $20\%$ at fixed $\bar{F}_{\rm cl}$. This is because the flux, across a wide range of wavelengths, is dominated by the same brightest stars, which correspond to the massive end of the IMF. Varying the IMF slope at fixed $\bar{F}_{\rm cl}$ changes the relative abundance between high-mass and low-mass stars, but hardly changes the absolute number of the former. Placing a significant constraint on the IMF slope is possible but will necessarily require a fairly precise determination of $\epsilon_2$.

Similarly, the right panel of \reffig{eps2} shows that for young populations $t_{\rm SSP} \lesssim 6\,$Myr, $\epsilon_2$ is mildly sensitive to the metallicity within the range $0.1< Z/Z_\odot < 1$, except in the case of very low metallicity $Z=0.1\,Z_\odot$ and through in bluest filter F275W sensitive to ionizing radiation, for which the stellar model predicts that $\epsilon_2$ in this filter can increase by a factor of two for $t_{\rm SSP}\gtrsim 3\,$Myr.  We also note that $\epsilon_2$ should be independent of dust reddening if all member stars are subject to homogeneous extinction.

Using the mass-magnification relation plotted in Figure 6 of \cite{vanzella2020ionizing}, we predict for the LyC Knot with $t_{\rm SSP}=3\,$Myr and $Z=0.6\,Z_\odot$ that $\epsilon_2 = (\mu_{\rm macro}/20)^{1/2}\,(0.5\textrm{--}0.6)\%$, insensitive to the high-mass IMF slope and essentially in all filters we consider here. 

The source stellar population can be reconstructed based on photometric and spectroscopic data. On the other hand, the value of $\epsilon_2$ could be inferred (using \refeq{rmsflux}) from combining a direct measurement of $\langle \Delta^2 \rangle_c$ and a theoretical prediction of $\langle \delta^2 \rangle_c$ through modeling the macro lens and the intervening microlens population. We could then pin down the absolute macro magnification $\mu_{\rm macro}$ needed to to have the right value of $\epsilon_2$ for the right source stellar population, which would allow us to determine the intrinsic mass and size of the star cluster.

Since multiple macro lensed images of the LyC Knot have the same $\epsilon_2$, one could directly measure $\langle \Delta^2 \rangle_c$'s between the various images, and translate that into the relative sizes of $\langle \delta^2 \rangle_c$. On the other hand, the relative $\mu_{\rm macro}$'s can be directly measured from photometry, and the signs can be inferred from the observed topology of the lensed image configuration. (Note that in the case of lensed images along a highly magnified, tangential arc with $|\mu_t| \gg |\mu_r|$, $\mu_r$ can be more easily extracted from a macro lens model without large uncertainties.) Since $\langle \delta^2 \rangle_c$ has nonlinear dependence on $\mu_{\rm macro}$, and the microlens population typically differs from one line of sight to another, we might be able to determine the correct absolute scale for the $\mu_{\rm macro}$'s without any accurate knowledge of $\epsilon_2$.

\subsection{Microlens Population of G1}

Having studied the intrinsic flux distribution of source stars in the LyC Knot, we now quantify flux (de-)magnification acting on individual stars, which is another ingredient in modeling the overall variability of the star cluster. We first need to characterize the stellar population of the microlens host galaxy G1.

As presented in \refapp{G1}, we perform SED fitting of the rest-frame UV continuum using the archival MUSE IFU data (Program 297.A-5012; PI: N. Aghanim). Our SED fitting procedures disentangle the light of G1 from the light of the lensed Images of the LyC Knot, and reveal that G1 has a spectroscopic redshift $z=0.458$, slightly higher than that of PSZ1-G311. As far as lensing of sources at $z_s=2.37$ is concerned, however, it is an excellent approximation to simply set the redshift of G1 to be the same as that of the galaxy cluster. 

Utilizing synthetic template SEDs from the \textsc{BPASS} model~\citep{eldridge2017binary, stanway2018re}, we determine that the stellar population of G1 has an age $t_{\rm SSP} = 10^{8.7\mbox{--}8.8}\,$yr, a metallicity in the range $0.4\lesssim Z/Z_\odot \lesssim 0.7$, and probably some small amount of internal dust extinction $E_{\rm G1}(B-V) \simeq 0.05$. Based on these parameters, we determine that the average surface mass density of microlenses amounts to $\kappa_\star = 0.007\mbox{--}0.02$, accounting for the fact that the line of sight toward Image 5 intersects the outskirt of G1.

In a preliminary lens model applicable to the region of Arc 1 as presented in \cite{rivera2019gravitational}, Image 5 is on the interior of the lensing critical curve, and hence its macro magnification factor is negative $\mu_{\rm macro}<0$. The absolute value $|\mu_{\rm macro}|$ is not precisely known. A conservatively estimated range would be $10<|\mu_{\rm macro}|<100$. Ongoing efforts to construct a detailed lens model for the Sunburst Arc~\citep{vanzella2020transient} may provide a more reliable number. The bottom line is that the expected large value of $|\mu_{\rm macro}|$, together with the estimate $\kappa_\star = 0.007\mbox{--}0.02$ from G1, indicates that $\kappa_\star\,|\mu_{\rm macro}| \sim \mathcal{O}(0.1$--$1)$. Therefore, Image 5 could be subject to optically thick statistical microlensing due to the intervening G1. Moreover, the negative sign of $\mu_{\rm macro}$ favorably enhances the microlensing effects on the flux compared to the $\mu_{\rm macro} > 0$ case~\citep{schechter2002quasar}.

\subsection{Magnification PDF}

\begin{figure*}
	\includegraphics[width=\textwidth]{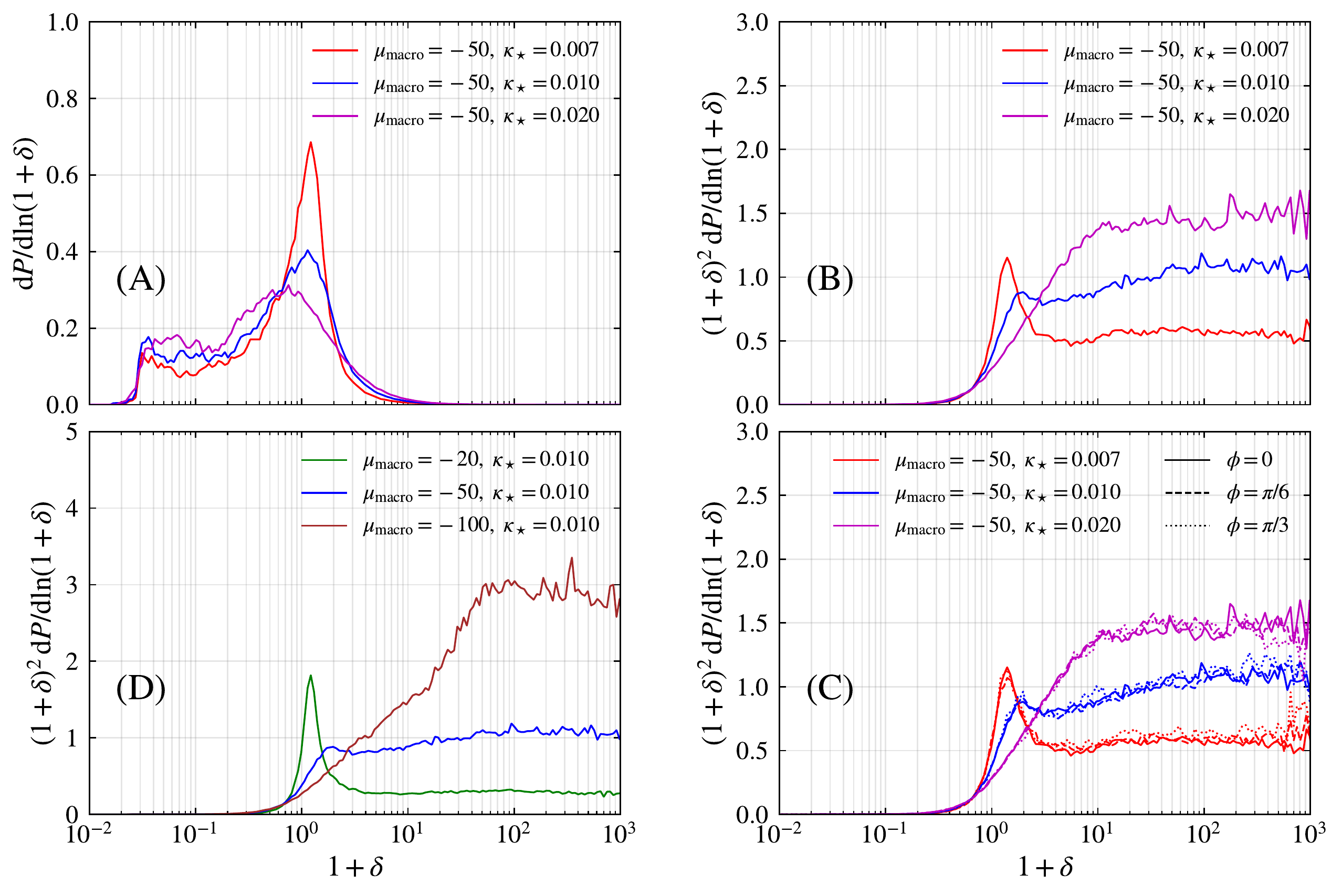}
    \caption{Probability distributions of the fractional variable flux $(1+\delta)$ induced by stochastic microlensing, for any individual source star residing in the LyC Knot. The distributions are derived, from randomly generated microlensing realizations, for Image 5 of the LyC Knot behind the foreground galaxy G1, for $\kappa_0=0.6$ and different values of $\mu_{\rm macro}$. The microlens mass function follows from an SSP of an age $t_{\rm SSP}=630\,$Myr and metallicity $Z=0.4\,Z_\odot$ with the standard IMF ($\alpha=2.3$), as inferred for G1 (\refapp{G1}). Except otherwise noted in Panel (C), the source trajectory is parallel to the macro shear $\phi = 0$. {\bf Panel (A)}: Probability density function (PDF) $\rmd P/\rmd \ln(1+\delta)$ for $\mu_{\rm macro}=-50$ and several choices for the mean convergence of microlenses $\kappa_\star=0.007,\,0.01,\,0.02$. While having the most support around $\delta = 0$, the PDF has a power-law tail $\rmd P/\rmd \ln(1+\delta) \propto (1+\delta)^{-2}$ for $\delta \gg 1$, and is only truncated by the effect of finite source size (typically at $1+\delta \simeq 10^4\,(R_s/10\,R_\odot)^{1/2}$ where $R_s$ is the stellar radius~\citep{2017ApJ...850...49V}). {\bf Panel (B)}: Same as (A), but with curves multiplied by $(1+\delta)^2$ to highlight the contribution to the second cumulant $\langle \delta^2 \rangle_c$. The power-law tail at $\delta \gg 1$ can make a dominant contribution to $\langle \delta^2 \rangle_c$ if $\kappa_\star \gtrsim 1/|\mu_{\rm macro}|$, which grows logarithmically with the (physical) cutoff in $\delta$. {\bf Panel (C)}: Same as (B), but with additional curves overplotted showing that the PDF is insensitive to the angle $\phi$ between the source trajectory and the direction of the macro shear. {\bf Panel (D)}: PDFs for at a fixed microlens abundance $\kappa_\star=0.01$, but with $\mu_{\rm macro}=-20,\,-50,\,-100$. The jaggedness of the curves in all panels, most noticeable for $\delta \gg 1$, arise from Monte-Carlo sampling noise.}
    \label{fig:mupdfs}
\end{figure*}

The PDF $P(\delta)$ for the flux fluctuation of any single source star needs to be calculated for given microlens mass function and surface density $\kappa_\star$, and at given macro magnification parameters $\mu_t$ and $\mu_r$. Previously, much efforts have been made to analytically model it in the study of microlensing induced variability of lensed quasars~\citep{1986ApJ...306....2K, PhysRevLett.59.2814, wyithe2001determining}. A number of convenient approximations have been found to work successfully in the low optical depth regime~\citep{peacock1982gravitational, 1983ApJ...267..488V, 1991A&A...251..393M, KofmanEtAl1997TwoD, LeeEtAl1997ThreeD}. For general values of $\kappa_\star$, however, accurate closed-form expressions for $P(\delta)$ that are fully satisfactory are not known. The major difficulty has to do with the nonlinear coupling between multiple microlenses at sufficiently high values of $\kappa_\star$ and the formation of many micro-images~\citep{fleury2020simple}. With an interest in quasar microlensing, some past studies presented numerical results for moderate values of $\mu_{\rm macro}$ (e.g.~\cite{GeraintIrwin1995AmpPDF}).

A task more analytically tractable than evaluating the full PDF $P(\delta)$ is to compute only the second cumulant $\langle \delta^2 \rangle_c$, or more generally the two-point correlation function at nonzero lags. In the literature, semi-analytic methods have been developed to avoid direct numerical simulation of random microlensing realizations~\citep{PhysRevLett.59.2814, Seitz1994microlensingI, Seitz1994microlensingII, Neindorf2003MicrolensingAutoCorr, Tuntsov2004ClusterMicrolensing}. However, the presented numerical results in those references were for either $\kappa_\star \sim \mathcal{O}(1)$ or $|1-\kappa_\star|\approx 0$, but are not easily quoted for the case under study in this work, i.e. a small mean convergence for the microlens $\kappa_\star=\mathcal{O}(10^{-2})$ coupled to a large macro magnification $|\mu_{\rm macro}|$. 

We therefore resort to numerical methods for evaluating $P(\delta)$. One class of methods generate randomized magnification patterns on the source plane through inverse ray-shooting~\citep{wambsganss1999gravitational}. For example, this was recently employed to quantify the statistics of microlensing on strongly lensed supernovae (SNe)~\citep{goldstein2018precise}. 

However, inverse ray-tracing is computationally costly for adequately resolving very large magnifications $\delta \gtrsim 10^2$ near micro-caustics, a regime crucial for the more compact stellar sources than for quasars and SNe. As we have argued before, the large-$\delta$ tail makes a significant or even dominant contribution to $\langle \delta^2 \rangle_c$. 

Instead, we implement a method proposed by \cite{lewis1993microlensing} (see also \cite{Witt1993EfficientMethod, GeraintIrwin1995AmpPDF}), which solves the continuous motion and the possible creation and annihilation of the micro-images of a moving point source. Based on solving ordinary differential equations, this method can find all micro-images and accurately track them to arbitrarily high magnifications~\citep{2017ApJ...850...49V}.

We first generate a random realization of microlenses in the presence of macro convergence and shear, and then solve the light curve for a point source that traverses a few Einstein lengths on the source plane, summing over all micro-images. By computing for a large number of random realizations, we build up a statistically adequate collection of random light curves, which enable us to evaluate $P(\delta)$ and $P_\delta(f)$, or any other statistics. While we numerically solve for point sources, we manually truncate the PDF at $\delta = 10^4$, appropriate for the typical stellar radii of massive main sequence stars or supergiants~\citep{2017ApJ...850...49V}.

In \reffig{mupdfs}, we present $P(\delta)$ numerically evaluated for the parameter ranges appropriate for Image 5, $0.007< \kappa_\star < 0.02$ and $-100<\mu_{\rm macro} < -20$. Qualitatively, most probability comes from a central core at $\delta \approx 0$, and some significant probability is associated with the strongly de-magnified ($1+\delta \ll 1$) regions expected for a negative $\mu_{\rm macro}$~\citep{chang1984star, schechter2002quasar, 2018ApJ...857...25D}. However, in nearly all cases what matters for the variance of flux fluctuation $\langle \delta^2 \rangle_c$ is primarily the $\delta \gg 1$ tail which nevertheless has very small weight in terms of probability. When $(1+\delta)^2\,\rmd P/\rmd \ln(1+\delta)$ is plotted, this clearly shows as a plateau for $\delta >0$, which confirms the power law $P(\delta) \propto \delta^{-3}$ and supposedly extends all the way to the cutoff magnification. As can be seen from \reffig{mupdfs}, the height of the plateau being comparable to the height of the $\delta \approx 0$ core marks the transition between optically thin and optically thick microlensing. \reffig{mupdfs} also shows that for fixed $\mu_{\rm macro}$ and $\kappa_\star$, the full PDF $P(\delta)$ and hence the value of $\langle \delta^2 \rangle_c$ are insensitive to the direction of source motion $\phi$.

For $\mu_{\rm macro}=-20$ and reliably extrapolating to a truncation at $\delta_{\rm max}=10^4$ following the $P(\delta) \propto \delta^{-3}$ law, we find $\langle \delta^2 \rangle_c = 1.6$, $2.5$ and $6.4$, for $\kappa_\star=0.007$, $0.01$ and $0.02$, respectively. If $\mu_{\rm macro}=-50$, those numbers become $\langle \delta^2 \rangle_c = 4.6$, $8.2$ and $11.4$. In this case, the RMS of the integrated flux variability $\Delta$ (defined as $\langle \Delta^2 \rangle_c$ in \refeq{rmsflux}) for Image 5 is enhanced by at least a factor $\sqrt{\langle \delta^2 \rangle_c} = 2$--$3$ relative to $\epsilon_2$, reaching $2\mbox{--}3\%$.

\subsection{Power Spectrum}

\begin{figure}
	\includegraphics[width=\columnwidth]{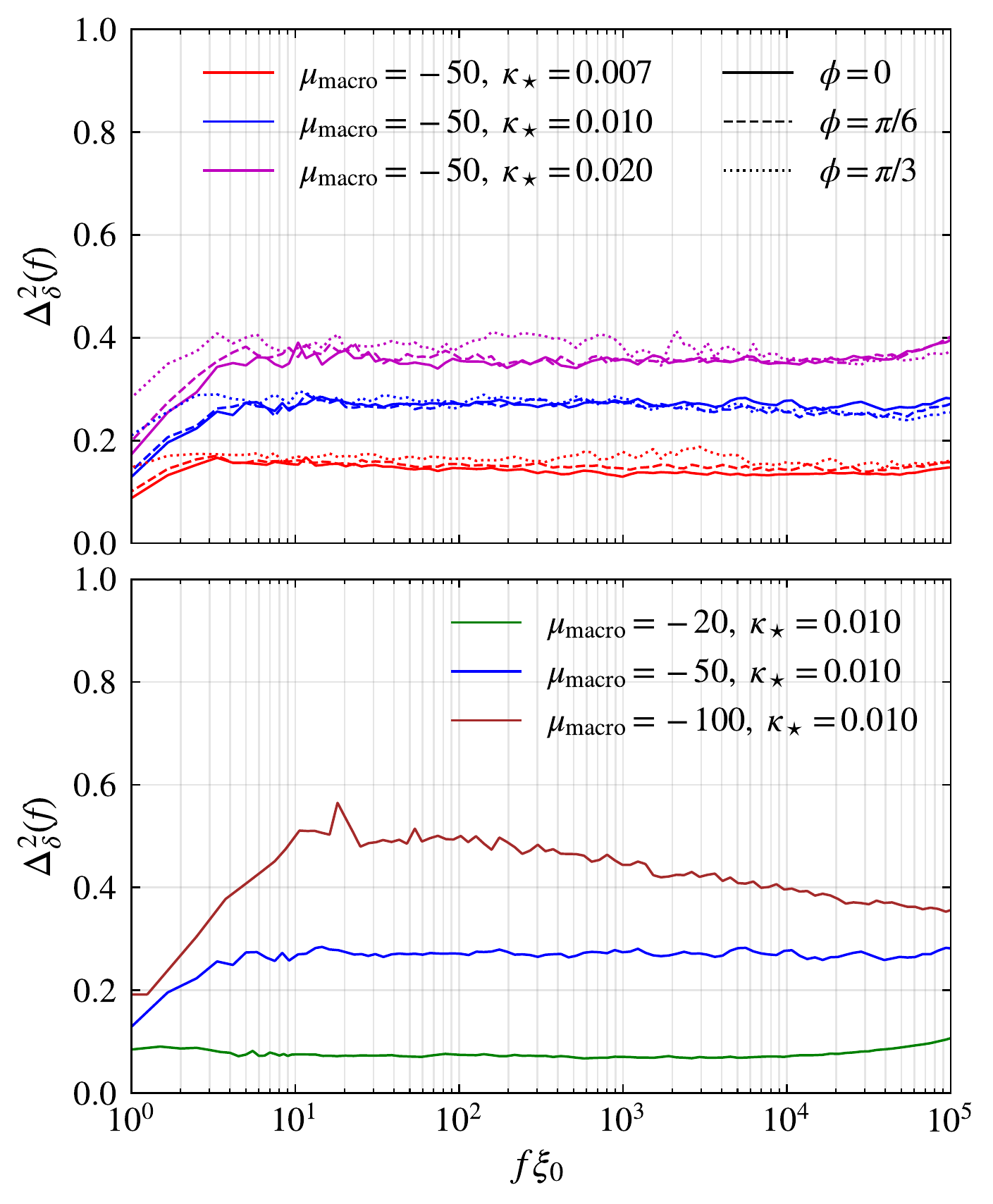}
    \caption{Characteristic power $\Delta^2_\delta(f)=f\,P_\delta(f)$ for the microlensing induced flux fluctuations of individual source stars, plotted as a function of the inverse spatial scale $f$ on the source plane. Curves have been computed for the inferred microlens population in the foreground galaxy G1. For the horizontal axis, $f$ is normalized to the inverse of the Einstein scale $\xi_0$ for a fiducial point microlens of $1\,\Msun$, which is about $2600\,$AU in the case of the LyC Knot, or equivalent to a timescale $\tau_0 \approx 40\,{\rm yr}\,(v_t/300\,{\rm km/s})^{-1}$ in the observer frame where $v_t$ is the effective transverse velocity \citep[Eq.~(12) of][]{2017ApJ...850...49V} of the LyC Knot relative to the lens. Most stellar microlenses in G1 have masses $\sim 0.2$--$0.3\,\Msun$. {\bf Top panel}: Cases of a fixed $\mu_{\rm macro}=-50$ but with increasing values of the microlens surface density $\kappa_\star=0.007,\,0.01,\,0.02$. We plot for different source motion directions across the magnification pattern on the source plane, $\phi=0$ (solid), $\pi/6$ (dashed), and $\pi/3$ (dotted). {\bf Bottom panel}: Cases of a fixed $\kappa_\star=0.01$ but with different macro magnification factors $\mu_{\rm macro}=-20$, $-50$, and $-100$ for Image 5. Except for the case of $\mu_{\rm macro}=-100$ and $\kappa_\star=0.01$ for which highly entangled micro-caustics are expected, $\Delta^2_\delta(f)$ exhibits a scale-invariant slope on scales well below the Einstein scales of the typical microlenses ($f \xi_0 > 10$). The trend that the curves start to rise near the shortest length scales (or the largest values of $f$) is likely due to artificial power induced by numerical interpolation of the simulated light curves near caustic crossing spikes.}
    \label{fig:mupowerspec}
\end{figure}

To understand the temporal structure of the flux variability, in \reffig{mupowerspec} we present the variability power spectrum $P_\delta(f)$ for an individual source star, derived from simulated light curves. Since the variability power spectrum $P_\Delta(f)$ of the entire star cluster is proportional to $P_\delta(f)$, information on the same timescales can be extracted by measuring $P_\Delta(f)$.

We measure the spatial frequency $f$ in units of $\xi^{-1}_0$, where $\xi_0$ is a fiducial Einstein scale
\begin{align}
    \xi_0 = \left( \frac{4\,G\,\Msun}{c^2}\,\frac{D_L\,D_{LS}}{D_S} \right)^{1/2},
\end{align}
corresponding to a $1\,\Msun$ microlens. In the case of microlensing toward Image 5, $\xi_0\approx 2600\,$AU. For the majority of the microlenses, which are sub-solar main-sequence stars of masses $\sim 0.2$--$0.3\,\Msun$, the Einstein scale is roughly half of $\xi_0$. 

Limited by computational cost, we only solve for light curve segments of finite lengths following the source motion across about order unity times $\xi_0$, resolving no more than a dozen micro-caustic crossing features per light curve for the values of $\mu_{\rm macro}$ and $\kappa_\star$ we consider. We thus do not have information of $P_\delta(f)$ for $f \xi_0 \ll 1$, but we do have reliable results for $f \xi_0 \gg 1$ thanks to the fine resolution of our adaptively sampled light curves. 

As shown in \reffig{mupowerspec}, $P_\delta(f)$ has a non-trivial shape around the characteristic Einstein scale of the microlenses, which should depend on the detailed microlens mass function. On the other hand, well below the Einstein scale $f \xi_0 \gg 1$, $P_\delta(f)$ shows a universal scale-invariant slope $P_\delta(f) \propto f^{-1}$, until truncated at $f R_S \sim \mathcal{O}(1)$ for a finite source radius $R_S$. This scale invariant power spectrum is a direct consequence of the universal behavior in the vicinity of a fold caustic, that $|\mu| \propto s^{-1/2}$ where $s$ is the distance to the caustic on the source plane~\citep{blandford1986fermat}, and therefore its Fourier transform goes as $f^{-1/2}$. We note that for $\mu_{\rm macro}=-100$ and $\kappa_\star=0.01$, our numerically evaluated $P_\delta(f)$ shows departure from perfect scale-invariance for moderately large values of $f \xi_0$. This probably can be attributed to the fact that at a very high number density micro-caustics strongly interfere with one another, which leads to violation of the simple scaling $|\mu| \propto s^{-1/2}$. 

The scale invariance of $P_\delta(f)$ for $f\xi_0\gg 1$ leads to an interesting conclusion: since $P_\Delta(f)$ is proportional to $P_\delta(f)$, and because the flux variability of the star cluster $\Delta$ is efficiently Gaussianized from the superposition of a large number of contributing source stars, $\Delta$ follows a scale-invariant process on those timescales.

In the case of the LyC Knot of Sunburst Arc, and assuming a typical observer-frame relative transverse velocity $v_t \sim 300\,{\rm km/s}$ between the source and the lens (see Eq.~(12) of \cite{2017ApJ...850...49V}), the source-plane Einstein scales for typical microlenses translate to a rather long timescale on the order of decades. This is unappealing that accumulating adequate statistics to make inference about $P_\delta(f)$ around the Einstein scale would probably be prohibitive from the observational perspective. On the other hand, we suggest that it would be much more feasible to measure the scale-invariant portion of the power spectrum, e.g. at $f \xi_0 \simeq 10^2$--$10^4$, which correspond to monitoring the LyC Knot over much more accessible timescales ranging from days to a year.

\subsection{Light Curve Example}

\begin{figure*}
	\includegraphics[width=\textwidth]{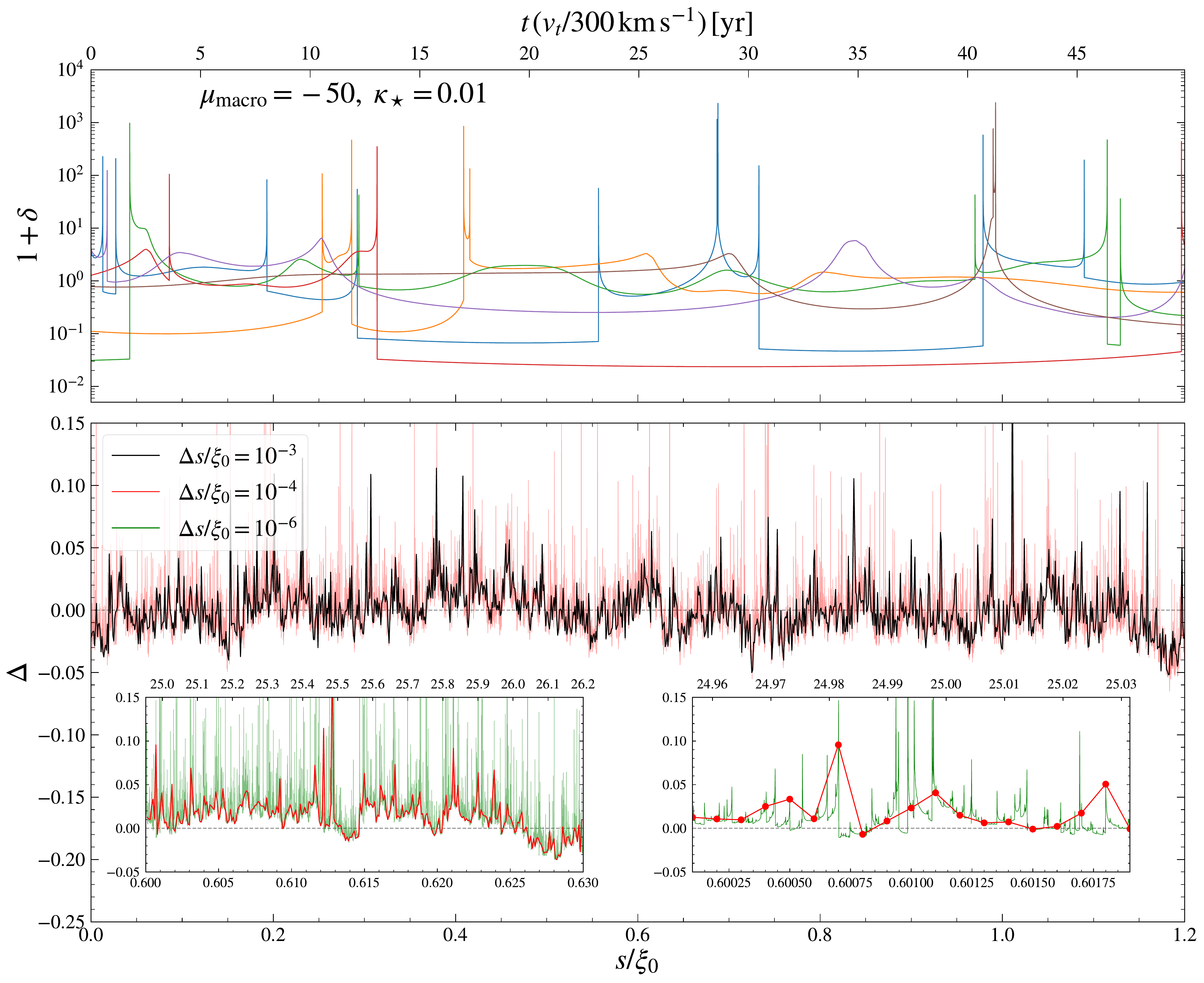}
    \caption{Example light curve of a mock star cluster that comprises the stellar population of the LyC Knot and has a total mass $M_\star=10^7\,\Msun$, assuming $\mu_{\rm macro}=-50$ and $\kappa_\star=0.01$. Because of computational limitation, we only explicitly evaluate microlensing light curves for about $34000$ brightest stars, which account for $90\%$ of the mean integrated flux and $99.8\%$ of the flux variance, but neglect the variabilities of the other fainter members. The bottom axis is the source-plane distance in units of the unit-solar-mass Einstein scale $\xi_0$, and in the top axis this is converted into a total elapsed time of about $50\,$yr for a fiducial $v_t=300\,{\rm km/s}$. We show the light curve in the F814W filter, but light curves in other filters from UV to near-IR are highly correlated. {\bf Top panel}: Random light curves for several individual stars (in different colors) uniformly sampled at $\Delta s = 10^{-6}\,\xi_0$. Individual stars brighten by orders of magnitude during micro-caustic transits, which nevertheless occur infrequently over the timescale of decades. {\bf Bottom panel}: Variability for the integrated flux uniformly sampled at different cadences $\Delta s/\xi_0 = 10^{-4}$ (red) and $10^{-3}$ (black). The {\bf left and right insets} zoom into short-timescale details, over a one-year timescale (left inset) and a one-month timescale (right inset) respectively, which are difficult to discern in the plot of the full light curve. For the insets, two cadences $\Delta s/\xi_0 = 10^{-4}$ (red) and $10^{-6}$ (green) are shown. The integrated flux exhibits the usual highly non-Gaussian flaring behavior on very short timescales (green), increasingly Gaussianizes on intermediate timescales (red), and approaches a Gaussian process on long timescales (black).}
    \label{fig:lightcurve_example}
\end{figure*}

\begin{figure}
	\includegraphics[width=\columnwidth]{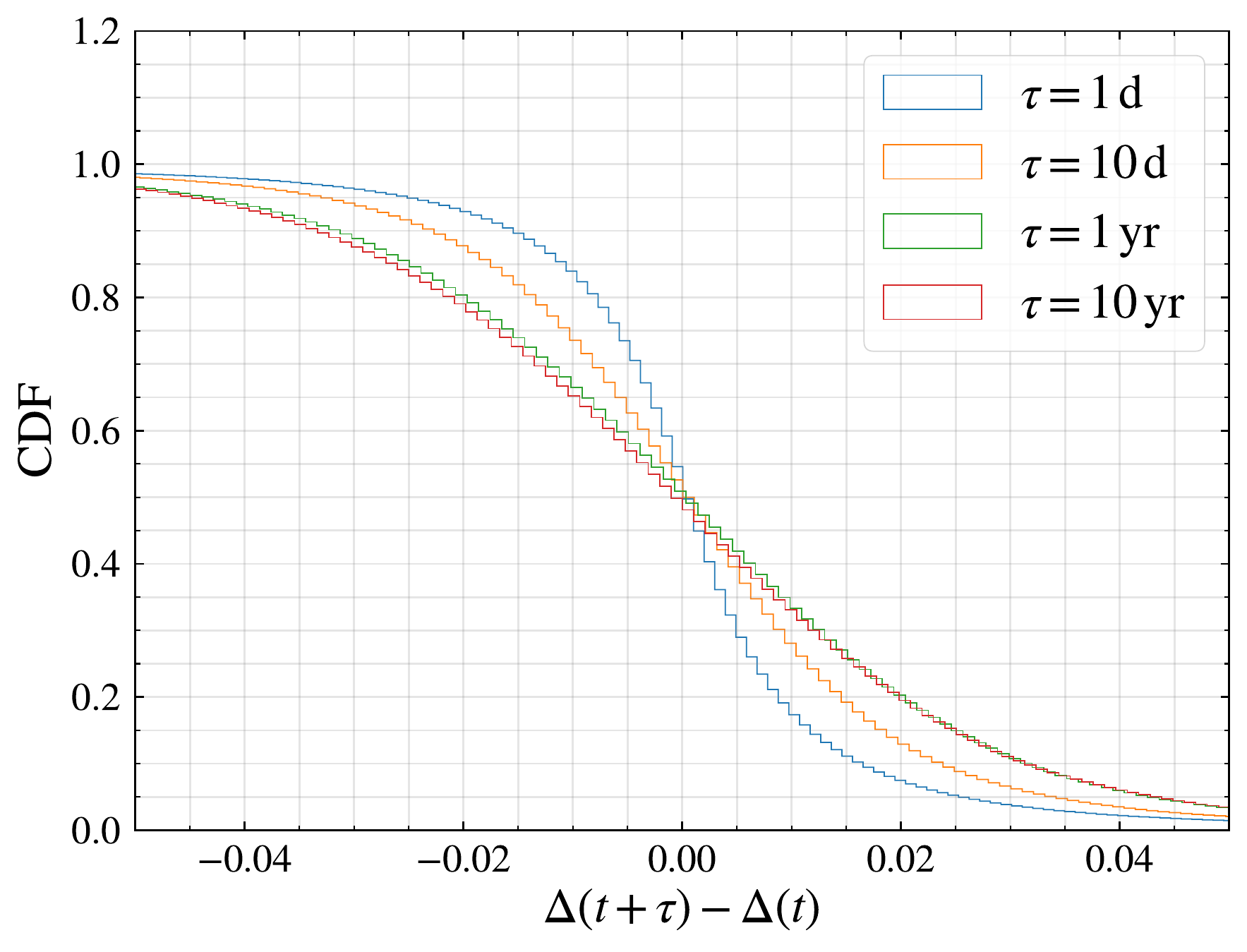}
    \caption{Cumulative distribution function of the fractional flux difference between two epochs separated by an amount of time $\tau$, derived from the simulated light curve in \reffig{lightcurve_example}. We show curves for several temporal separations, $\tau=1\,$d, $10\,$d, $1\,$yr and $10\,$yr.}
    \label{fig:DeltaCDF}
\end{figure}

To gain insight into the variable nature of the integrated flux, we show in \reffig{lightcurve_example} one example of a synthetic light curve for the LyC Knot. To be specific, we consider $\mu_{\rm macro}=-50$ and $\kappa_\star=0.01$, and set a cluster mass $M_\star=10^7\,\Msun$~\citep{vanzella2020ionizing}. It is computationally prohibitive to explicitly simulate all member stars~\citep{LewisIbataWyithe2000MACHO} as there are $\sim \mathcal{O}(10^7)$ of them! For a good approximation, we only include the flux variability from a subset consisting of $\sim 34000$ bright stars. This subset account for $90\%$ of the total mean flux $\bar{F}_{\rm cl}$ and $99.8\%$ of the variance parameter $\epsilon_2$ (and hence of $\langle \Delta^2 \rangle_c$). According to \refeq{Deltacumu}, higher order cumulants of the cluster variability, $\langle \Delta^n \rangle_c$ for $n=3,\,4,\,\cdots$, are increasingly dominated by the brightest member stars. We include the contribution to $\bar{F}_{\rm cl}$ from the numerous other fainter stars but simply neglect their variabilities. We note however that during micro-caustic crossings many of the faint stars can be magnified by larger factors than the brighter stars due to their small stellar sizes.

For an individual star, the flux evolves slowly for most of the time. Occasionally, the flux rapidly increases by several orders of magnitude around micro-caustic crossings, but (e.g. for $\mu_{\rm macro}=-50$ and $\kappa_\star=0.01$) such events only occur infrequently over decades. For $\mu_{\rm macro}<0$, the flux can hide for a long period of time, but in this case the fractional flux change is bounded $\delta > -1$.

By contrast, the integrated flux reflects the collective effect of many source stars. As shown in the lower panel of \reffig{lightcurve_example}, on long timescales $f \xi_0 \approx 1$--$100$, i.e. from years to decades, the light curve resembles a scale-invariant Gaussian process, which, for the case $\mu_{\rm macro}<0$ we consider, is mainly driven by many source stars independently entering or leaving the extensive regions of low magnification in between caustic pairs~\citep{chang1984star, 2018NatAs...2..334K, 2018ApJ...857...25D}. On the other hand, when sampled at high resolutions, i.e. $f \xi_0 \approx 100$--$ 10^4$ on timescales from days to months, the variable flux exhibits a highly non-Gaussian behavior. The most prominent feature of this behavior are flares from micro-caustic crossings that occur intermittently and frequently (referred to as the ``extreme events'' by \cite{LewisIbata2000MACHO}; also note that the fast transients reported by \cite{2018NatAs...2..324R} may be of such nature). We further note that on short timescales our simplification of including only the brightest stars must have left out a large number of lower-amplitude flaring events from micro-caustic crossings of the many more fainter stars. 


Numerically, we find an overall $\sqrt{\langle \Delta^2 \rangle_c}\approx 2.5\%$ (corresponding to the green curve in \reffig{lightcurve_example} with the finest temporal sampling $\Delta s/\xi_0 = 10^{-6}$), in good agreement with \refeq{rmsflux} and \refeq{eps2}. \reffig{DeltaCDF} shows the probability distribution for the fractional flux difference between two random epochs separated by an amount of time $\tau$, i.e. $\Delta(t+\tau) - \Delta(t)$, computed using the simulated light curve shown in \reffig{lightcurve_example}. Even at $\tau=1\,$d, the typical values for $|\Delta(t+\tau) - \Delta(t)|$ can be $\sim 0.5\%$, and this number grows to $\sim 1$--$2\%$ as the temporal baseline increases to one year. The distribution stays nearly the same between one-year and one-decade separations, except for extreme values of $|\Delta(t+\tau) - \Delta(t)|$. These results regarding the temporal structure indicate that the timescales of observational baselines must be taken into account when flux variability measurements at multiple epochs are compared to theory, and also have implications for the optimal observing strategy.

\begin{figure*}
    \hspace{-1cm}
	\includegraphics[width=\textwidth]{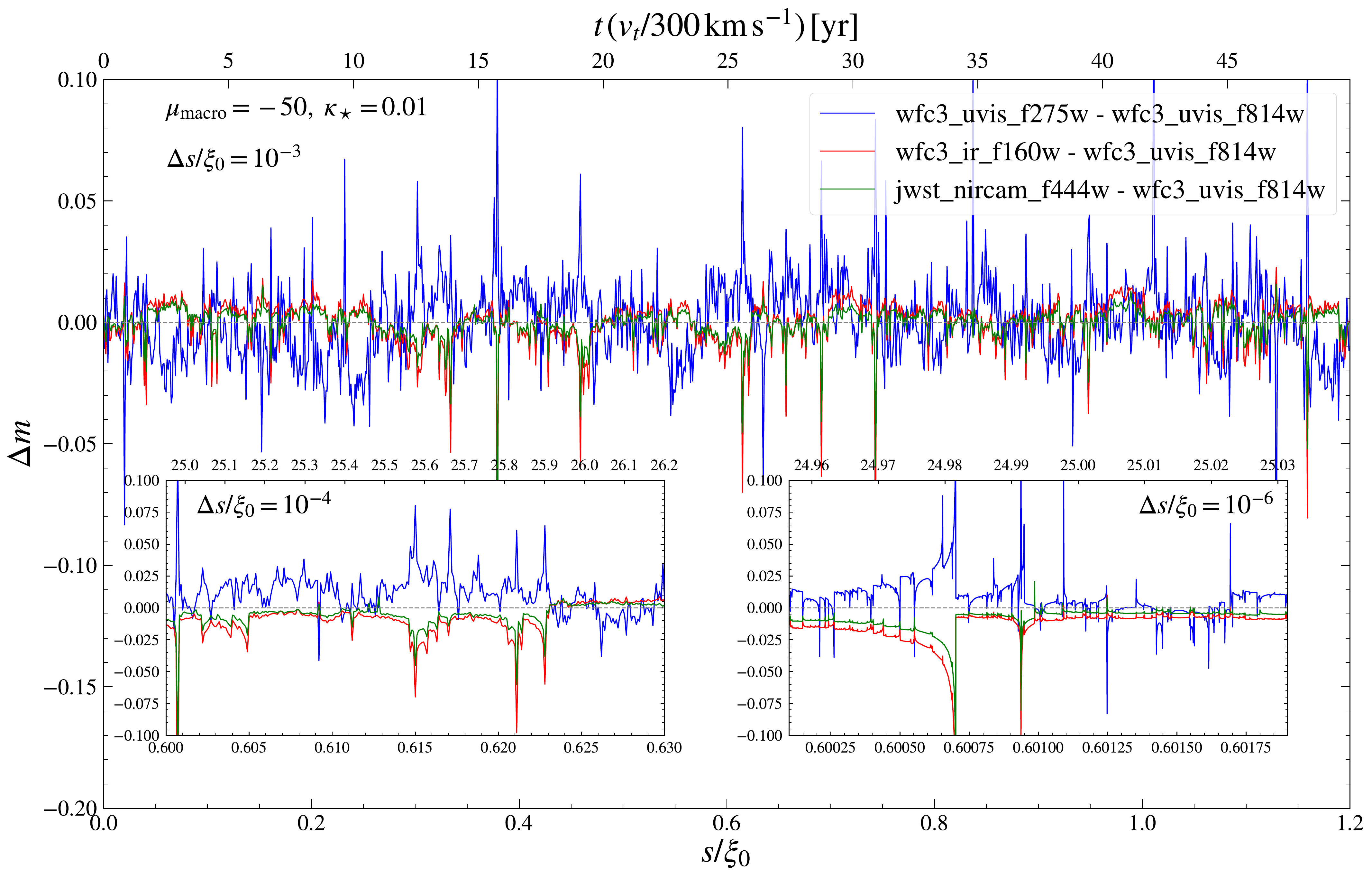}
    \caption{The same random light curve realization as in \reffig{lightcurve_example} but showing the color variation measured in magnitudes. Curves are uniformly sampled at a coarse resolution $\Delta s = 10^{-3}\,\xi_0$ in the {\bf main panel}, and are sampled at finer resolutions at $\Delta s = 10^{-4}\,\xi_0$ and $10^{-6}\,\xi_0$ in the {\bf left and right insets}, respectively. On long timescales, the color variability resembles a Gaussian random process. On short timescales, rapid events of large amplitude color variation are induced when individual bright stars transit micro-caustics.}
    \label{fig:lightcurve_example_color}
\end{figure*}

Another inevitable consequence is the random variation of the apparent color of the star cluster, defined as the magnitude difference between two chosen filters. This is examined in \reffig{lightcurve_example_color}, which is made based on the same realization used in \reffig{lightcurve_example}. Measured relative to the HST F814W filter as the reference filter, we show the colors for the HST F275W filter, which uniquely probes hydrogen ionizing radiation, that for the HST F160W filter, which measures the rest-frame optical flux, and that for the JWST NIRCAM F444W filter, which measures the rest-frame near-IR flux. On timescales of years to decades, the color variabilities follow a Gaussian random behavior, with amplitudes on par with that of the flux variability. Similarly, non-Gaussian transient events of large-amplitude color variation are expected to be seen on shorter timescales, which are mainly associated with the most luminous individual stars transiting micro-caustics.

\section{Discussion}
\label{sec:disc}

In this section, we comment on two issues: confusion with intrinsic variable stars, and stellar multiplicity. Furthermore, we suggest that the phenomenon of statistical microlensing can be used to constrain the abundance of compact objects that could partially constitute the DM.

\subsection{Intrinsically Variable Stars}
\label{sec:intrinsic_variable_star}

So far in our analysis we have assumed that all source stars have stationary fluxes in the absence of microlensing. This is an idealized assumption, as stars can exhibit variable fluxes, on a range of timescales (days to decades) and to various degrees, due to changes in the physical properties of the stars themselves.

When the microlensing optical depth is significant $\mathcal{O}(0.1$--$1)$, as we have studied before, every star is subject to order-unity flux variation due to lensing. Intrinsic variability can be a significant confusion effect in the context of this study only if the stars that dominate the observed total flux in a given band commonly have order-unity intrinsic variability. Observational data suggest that this is not the case for young stellar populations of only a few Myr old observed at short wavelengths. For example, \cite{Controy2018M51variability} monitored the Whirlpool Galaxy (M51) using the HST for stellar variability over the timescale of days to months. Complete to $M_{I,\,814}<-4$ in the $i$-band, that study is sensitive to a diversity of luminous stars of a wide range of ages. The study found that all stars with $M_{I,\,814}<-3$ have relatively small amplitudes of intrinsic flux variation, $\sqrt{\VEV{\delta^2}_c} < 0.45$ (Figure 11 of \cite{Controy2018M51variability}), except for Mira variables (Miras) and semi-regular variables (SRVs) which are typically red giant stars.

Our population model for the LyC Knot indicates that red stars with $M_{V,\,606} - M_{I,\,814} > 1$ have completely negligible contribution to F814W (rest-frame UV $\sim 2000$--$2500\,\AA$) for $t_{\rm age} < 50\,$Myr, and only have a sub-percent contribution to the total flux in F160W (rest-frame optical $\sim 5000\,\AA$) even at 50 Myr. Therefore, pulsating red giant stars such as Miras and SRVs are unimportant for the study of very young star clusters (a few Myr) at rest-frame UV or optical wavelength. Neglecting these red variables, all other luminous stars with $M_{I,\,814}<-3$ contribute an upper limit $\sqrt{\langle\delta^2\rangle_c} < 0.45$, which, when applied to all stars, corresponds to only $\sqrt{\langle\Delta^2\rangle_c} < 0.45\,(\mu_{\rm macro}/20)^{1/2}\,0.6\% \simeq (\mu_{\rm macro}/20)^{1/2}\,0.3\%$ for the variability of the integrated flux. In fact, at 3 Myr all stars with $M_{I,\,814}<-3$ account for $90\%$ ($87\%$) of the mean integrated flux $\langle\bar{F}\rangle$ and $99.8\%$ ($99.9\%$) of the second moment $\langle\bar{F}^2\rangle$ in the F814W (F160W) filter. Additionally, this is likely to significantly overestimates the intrinsic variability, as \cite{Controy2018M51variability} reported that the bright stars with $M_{I,\,814}<-6$ only have $\sqrt{\langle\delta^2\rangle_c} < 0.1$. At 3 Myr, these stars contribute to 54\% (57\%) of $\langle \bar{F} \rangle$ and 90\% (94\%) of $\langle\bar{F}^2\rangle$ in the F814W (F160W) filter. Also, \cite{Controy2018M51variability} showed that the bluest stars on average have the lowest variability fraction, as those are mainly main sequence stars. Indeed, at 3 Myr we expect stars with $M_{V,\,606} - M_{I,\,814}<0.4$ account for nearly the entirety of the integrated flux in F814W and F160W. Hence, intrinsic stellar variability in comparison to microlensing is probably negligible for a young star cluster of only a few Myr old when observed at rest-frame UV or optical wavelengths, as hot main sequence stars are photometrically very stable~\citep{Laur2017variability}, and hot evolved (super-)giant stars (e.g. luminous blue variables~\citep{Smith2017LBVreview}) are rare in such a young cluster. For aged stellar systems at 100 Myr or older, or observation at rest-frame infrared wavelengths, pulsation or stellar outbursts of unstable evolved stars may cause confusion with microlensing induced flux variability.

Apart from time-dependent changes in the intrinsic properties of the stars, eclipsing binaries can exhibit variable fluxes of large amplitudes. What is relevant to the case of the LyC Knot is the eclipsing binary fraction for OB stars. Using HIPPARCOS data, \cite{Lefevre2009OBstarsHIPPARCOS} identified 127 eclipsing binaries out of $\sim 2400$ local OB stars. In particular, the fraction of eclipsing binaries with large flux variability amplitude $>0.4\,$mag is $\sim 1\,\%$. The generally small fraction of eclipsing binaries does not lead to any sizable flux variability in the cluster integrated flux $\bar{F}$ that is comparable to microlensing effects.

More dramatic events such as supernova explosions are rarer to be found in any single star cluster, and when such a event does happen it will show unique photometric and spectroscopic behaviors to be distinguished from microlensing effects.

We emphasize that it is possible to empirically distinguish between microlensing effects and intrinsic variability by observing multiple lensed images of the same star cluster as in the case of the LyC Knot. It is expected that some images (such as Image 5) are subject to significantly more foreground microlensing than the others. The flux variance observed for the latter can be used to set the base intrinsic variability, and any excess variability observed for the former is a robust signature of microlensing.

\subsection{Stellar Multiplicity}


when building up a synthetic stellar population using FSPS and computing $\epsilon_2$, we have neglected the possibility of two or more stars forming a tightly bound system. In reality, massive stars in young star clusters are found to commonly have significant companions~\citep{Sana2012MassiveBinaries, Sana2013MultipleOStars, Dunstall2015MultipleBstars}. This leads to an underestimate in $\epsilon_2$ because members of a multiple stellar system traverse the source-plane magnification pattern in a correlated way. For example, if every star has an equal companion with perfectly correlated variability, then $\epsilon_2$ increases by a factor $\sqrt{2}\approx 1.4$. This, in principle, opens up the interesting possibility of inferring the binary or multiple fraction of massive stars from the RMS amplitude of the flux variation. To obtain robust results, however, one must accurately model both the microlens and the source population. State-of-the-art modeling of the source population must be carried out using all photometric and spectroscopic data, accounting for uncertainty in IMF, metallicity, dust reddening, and so on. The results should be verified for a more general star-formation history than what is assumed here. Another dynamical effect, which is interesting in its own right, is that rejuvenated or exotic stars may efficiently form from stellar collisions in a dense star cluster~\citep{Gaburov2008collI, Gaburov2010collII}, which may significantly alter the stellar luminosity function and hence the RMS flux variation.

Those degeneracies may be broken by considering the imprints of binary or multiple stars in the variability power spectrum $P_\Delta(f)$. This is because, depending on the separation, the correlation in variability can be perfect on long timescales but degrades substantially on short timescales. Even very close companion stars still cross the same micro-caustic at slightly different times. This effect should break the scale-invariance of the power spectrum $P_\Delta(f)$ found on sub-Einstein scales (\reffig{mupowerspec}). Through a detailed measurement of the integrated flux variability, we may be able to infer not only the abundance of binary or multiple stars but also their separation distribution. More detailed investigations are left for future work.

\subsection{Probing Compact Dark Matter}

Extremely compact objects such as primordial black holes (PBHs)~\citep{carr1974black, meszaros1974behaviour, chapline1975cosmological} may constitute some significant fraction of the Dark Matter (DM) or make gravitational-wave sources detectable to LIGO/Virgo~\citep{nakamura1997gravitational, bird2016did, sasaki2016primordial}. Their abundance can be observationally probed based on their microlensing effects~\citep{paczynski1986gravitational, griest1991galactic}. Microlensing analyses of Galactic stars~\citep{alcock2001macho, griest2013new}, toward local-group galaxies~\citep{tisserand2007limits, 2019NatAs...3..524N}, and of extragalactic SNe~\citep{zumalacarregui2018limits}, stars~\citep{2018ApJ...857...25D, Oguri:2017ock} and quasars~\citep{mediavilla2017limits}, together with other astrophysical~\citep{Brandt2016MACHOconstraint} or cosmological tests~\citep{AliHaimoud2017machoCMB}, have constrained the mass fraction $f_{\rm PBH}$ of planetary or stellar mass PBHs to be no more than a few to tens of percent.

Flux variability induced by statistical microlensing acting on lensed star clusters would offer a novel opportunity to constrain the PBH abundance owing to a large DM column density along the line of sight. Indeed, it was proposed before that one could monitor surface brightness variations in galaxies at $z \sim 0.5$~\citep{LewisIbata2000MACHO} or in giant arcs at $z \sim 1$ magnified by galaxy clusters~\citep{LewisIbataWyithe2000MACHO, Tuntsov2004ClusterMicrolensing} in order to probe a PBH population of $f_{\rm PBH} = \mathcal{O}(1)$.

Regarding the Sunburst Arc, the multiple images of the LyC Knot other than Image 5 could serve this purpose because our ICL modeling suggests $\kappa_\star \lesssim 0.002$ at the location of the Sunburst Arc for normal stellar microlenses. Given a PBH mass function, one would neglect intracluster stars by setting $\kappa_\star = f_{\rm PBH}\,\kappa_0$, and repeat the same analysis as in the previous sections.

According to the discussion of \cite{2017ApJ...850...49V}, if PBHs in the mass range $\sim 10^{-6}\,\Msun$--$10^2\,\Msun$ contribute more than $f_{\rm PBH} \simeq 2\%$ of all DM, there will be an equivalent $\kappa_\star \gtrsim 0.01$, comparable to or exceed the (effectively) optically thick surface abundance of microlens stars from G1, and possibly a resultant fluctuating flux varying at the percent level. This suggests, through dedicated photometric monitoring of just one system, that statistical microlensing of magnified extragalactic star clusters has the potential to probe $f_{\rm PBH}\simeq \mathcal{O}(10^{-2})$ over a wide range of PBH masses. This would be highly competitive, and in particular complementary at sub-solar PBH masses, to other proposed methods based on microlensing of extragalactic radio bursts~\citep{Munoz:2016tmg, 2020arXiv200313349L}, GRBs~\citep{JiKovetzKamionkowski2018GRBMicrolensing} and chirping gravitational waves~\citep{PhysRevLett.122.041103}. We note, however, that a higher abundance of PBHs is not necessarily easier to probe; a value for $\kappa_\star = f_{\rm PBH}\,\kappa_0$ that is too large can instead reduce the fractional variability~\citep{Tuntsov2004ClusterMicrolensing}. The optimal value of $f_{\rm PBH}$ to probe therefore depends on $\mu_{\rm macro}$, which will be different toward different macro images of the same star cluster or different star clusters. In light of future access to applicable photometric data, further study is warranted to derive the precise amplitude of flux variability for a broad range of $f_{\rm PBH}$ and PBH mass distributions.

\section{Conclusion}
\label{sec:concl}

We have studied the microlensing-induced flux variability of gravitationally magnified star clusters, which are typically found in lensed star-forming galaxies at high redshifts. We have made the suggestion that in galaxy cluster strong lenses the condition for optically thick microlensing is often satisfied because the usually small surface density of foreground stellar microlenses $\kappa_\star = \mathcal{O}(10^{-2})$ is coupled to a large macro magnification $|\mu_{\rm macro}| = \mathcal{O}(10\mbox{--}10^2)$. Very young star clusters of only a few Myr old are particularly advantageous because their fluxes are dominated by a relatively small number of massive O main-sequence stars or evolved B supergiants~\citep{GilMerinoLewis2006StarFormingRegion}, and hence the flux variability does not average out completely.

We have used a statistical framework to quantify the flux variability. We have shown that the variable flux exhibits rapid, frequent flares of large amplitude which arise from micro-caustic transits by individual stars, typically on timescales from days to months, while has a scale-invariant, Gaussian behavior of small amplitude on longer timescales from years to decades. 
We have conducted a case study of the LyC Knot discovered in the Sunburst Arc behind the galaxy cluster lens PSZ1-G311. We found that a minor foreground galaxy intervenes the line of sight toward one magnified image of the LyC Knot, providing a significant surface density of stellar microlenses. Through a population synthesis study, we predict that this lensed image should exhibit flux variability at the $\sim 1$--$2\%$ level, across a wide wavelength range from rest-frame UV to near-IR, and depending on the precise value of its macro magnification $\mu_{\rm macro}$. This prediction only mildly depends on the IMF slope and metallicity. Measuring the variability statistics therefore offers a microlensing-based method to determine the macro magnification $\mu_{\rm macro}$, and to probe the fraction of stellar multiplicity in massive stars.

How feasible would it be to measure this light variability? Theoretically, an SNR of 100 is needed to have 1\% precision. Since the lensed images of the LyC Knot are remarkably bright ($22$--$23$ mag), $1$--$2\%$ photometry should be achievable with $\sim 1$--$3\,$ks exposures using the HST through observed optical or near-IR filters. With the forthcoming JWST, even shorter exposures will be sufficient. In fact, \cite{rivera2019gravitational} detected the lensed images at SNR$>10^3$ and reported aperture photometry precisions better than $0.1\%$ with a total $\sim 5\,$ks exposure in the HST F814W filter. Clearly, photon count noise will not be the major limitation from the space, in line with the earlier assessment of \cite{LewisIbata2000MACHO}. Even without adaptive optics, percent-level photometry should be feasible from the ground using 4-meter class telescopes if photometry is calibrated using reference stars. In either case, the key will be to uniform image reduction procedures and photometry schemes so as to prevent any systematic biases when comparing data acquired at different epochs, and to carefully subtract any uniform sky background that might differ between epochs. In this regard, applying the same analysis to lensed images unaffected by microlensing can serve as a control test. Ideally, measurements made at a large number of epochs will be required to enable a statistically meaningful comparison with the theory. It would be certainly an economic strategy to combine such a program with others that monitor the field of view of interesting galaxy cluster lenses in search of SNe and caustic transients.

Finally, we have pointed out that magnified images without significant intervention from foreground stars are clean targets to probe planetary to stellar mass compact objects in the galaxy cluster's DM halo. In the case of the Sunburst Arc, for example, we suggest that monitoring flux variability for the other lensed images of the LyC Knot will allow to probe as little as just a few percent of all DM in this form.  

\appendix

\section*{Acknowledgements}

The author would like to thank Matthew Bayliss, Geraint Lewis, Jordi Miralda-Escud\'{e}, Emil Rivera-Thorsen, Artem Tuntsov, and Keiichi Umetsu for useful comments. The author also thanks the anonymous referee for a constructive review of the manuscript. The author acknowledges the support of John Bahcall Fellowships at the Institute for Advanced Study. This work is based on observations made with the NASA/ESA Hubble Space Telescope, obtained from the data archive at the Space Telescope Science Institute (STScI). STScI is operated by the Association of Universities for Research in Astronomy, Inc. under NASA contract NAS 5-26555. For our analysis, we use archival HST images available from the Barbara A. Mikulski Archive for Space Telescopes\footnote{\url{https://mast.stsci.edu/portal/Mashup/Clients/Mast/Portal.html}}, taken with the Wide Field Camera 3 (WFC3) from three programs: (1) Proposal \# 15101 (PI: H. Dahle) using F555W, F814W, F105W and F140W; (2) Proposal \# 15377 (PI: M. Bayliss) using F606W, F125W and F160W; (3) Proposal \# 15418 (PI: H. Dahle) using F275W. 
This work is also based on spatially resolved spectroscopic data collected at the European Southern Observatory (ESO) under ESO programme 297.A-5012 (PI: N. Aghanim). We accessed the public VLT/MUSE data from the ESO Archive\footnote{\url{http://archive.eso.org/scienceportal/home}}.


\section*{Data Availability Statement}

The original data underlying this work are all publicly available from the Barbara A. Mikulski Archive for Space Telescopes and from the European Southern Observatory Archive. The derived data generated in this research will be shared upon reasonable request to the corresponding author.


\appendix



\section{Dark Matter Halo of PSZ1 G311.65-18.48 and Intracluster Light}
\label{app:clusterhalo}

The foreground lens, massive galaxy cluster PSZ1 G311.65-18.48 at $z_l=0.44$, was first catalogued by the all-sky {\it Planck} galaxy cluster survey~\citep{2014A&A...571A..29P} through the Sunyaev-Zeldovich (SZ) effect~\citep{1972CoASP...4..173S}. At the source redshift $z_s=2.37$ of the Sunburst Arc, it has a spectacular angular Einstein radius $\theta_{\rm E}\simeq 30\arcsec$~\citep{dahle2016discovery}.

The multiple sections of the Sunburst Arc trace an ideal Einstein ring, which suggests that the cluster lens is highly spherical and dynamically relaxed. Neglecting a possibly small ellipticity, the observed Einstein radius $\theta_{\rm E}$ translates into a total projected mass $M_{\rm E}=1.7 \times 10^{14}\,M_\odot$ enclosed within the Einstein radius. 

Assuming that the stellar and gaseous contributions to $M_{\rm E}$ are subdominant~\citep{rasia2004dynamical}, the observed Einstein scale can be reproduced by a spherical Navarro-Frenk-White (NFW) profile~\citep{1996ApJ...462..563N, navarro1997universal} with a concentration parameter $C_{200}=8$ (throughout overdensity is defined relative to the cosmic critical density), a dynamic mass $M_{200}=1.0\times 10^{15}\,M_\odot$, a scale radius $R_s = 224\,$kpc, and a lensing convergence at the Einstein radius $\kappa_{\rm E}=0.6$. This model gives $M_{500}=7.7 \times 10^{14}\,M_\odot$, in agreement with the SZ-based dynamic mass estimate $M_{500{
\rm SZ}}=6.6^{+0.9}_{-1.0}\times 10^{14}\,M_\odot$~\citep{dahle2016discovery}. 

The chosen concentration parameter $C_{200}=8$ is reasonable for high-mass ``superlens'' clusters~\citep{OguriBlandford2009LargestEinsteinRadii, Umetsu2020ClusterLensingReview} at $z \lesssim 0.5$ characterized by large Einstein radii $\theta_{\rm E} \gtrsim 30\arcsec$ (for $z_s \sim 2$) and $M_{200} \sim 2\times 10^{15}\,M_\odot$.
If instead a higher value $C_{200} =12$ is adopted, the profile more resembles an isothermal one. The dynamic mass is then reduced to $M_{200} = 7 \times 10^{14}\,M_\odot$, with a decreased scale radius $R_s = 133\,$kpc and a smaller Einstein convergence $\kappa_{\rm E}=0.5$. Even in this case, $M_{500}=5.6\times 10^{14}\,M_\odot$ is still fairly consistent with the SZ-based mass measurement. For both choices, $2\,(1-\kappa_{\rm E})=0.8$--$1.0$ is close to unity, which is consistent with the narrow appearance of the arc's width. We use the value $C_{200}=8$ for our fiducial cluster halo profile, according to which we use a fiducial value $\kappa_0=0.6$ throughout this work.

To determine the intracluster stellar population of PSZ1 G311, we perform SED fitting using the archival MUSE WFM spatially resolved spectroscopy data (Program 297.A-5012; PI: N. Aghanim). The data were acquired during May-August 2016 with seeing $0.5$--$0.8\arcsec$ and a total exposure time of 1.2 hr, and have been reduced using the standard MUSE data reduction pipeline. The SNR is insufficient for a direct detection of the ICL in the MUSE data out to the Einstein radius, so we instead perform SED fitting within an aperture nearer to the brightest cluster galaxy (BCG), using BPASS synthetic SED templates. Assuming an SSP with a standard IMF and subject to Milky-Way dust reddening, we find an age $t_{\rm SSP} \approx 5\,$Gyr and a sub-solar metallicity $Z=0.4\,Z_\odot$. We then apply this solution to the ICL surface brightness in HST wide filters at the Einstein radius, which we find is fainter than $25\,{\rm mag}/{\rm arcsec}^2$ in the F160W filter, and hence derive $\kappa_\star \lesssim 0.002$.




\section{Stellar population of the foreground galaxy G1}
\label{app:G1}

\begin{figure}
	\includegraphics[width=\columnwidth]{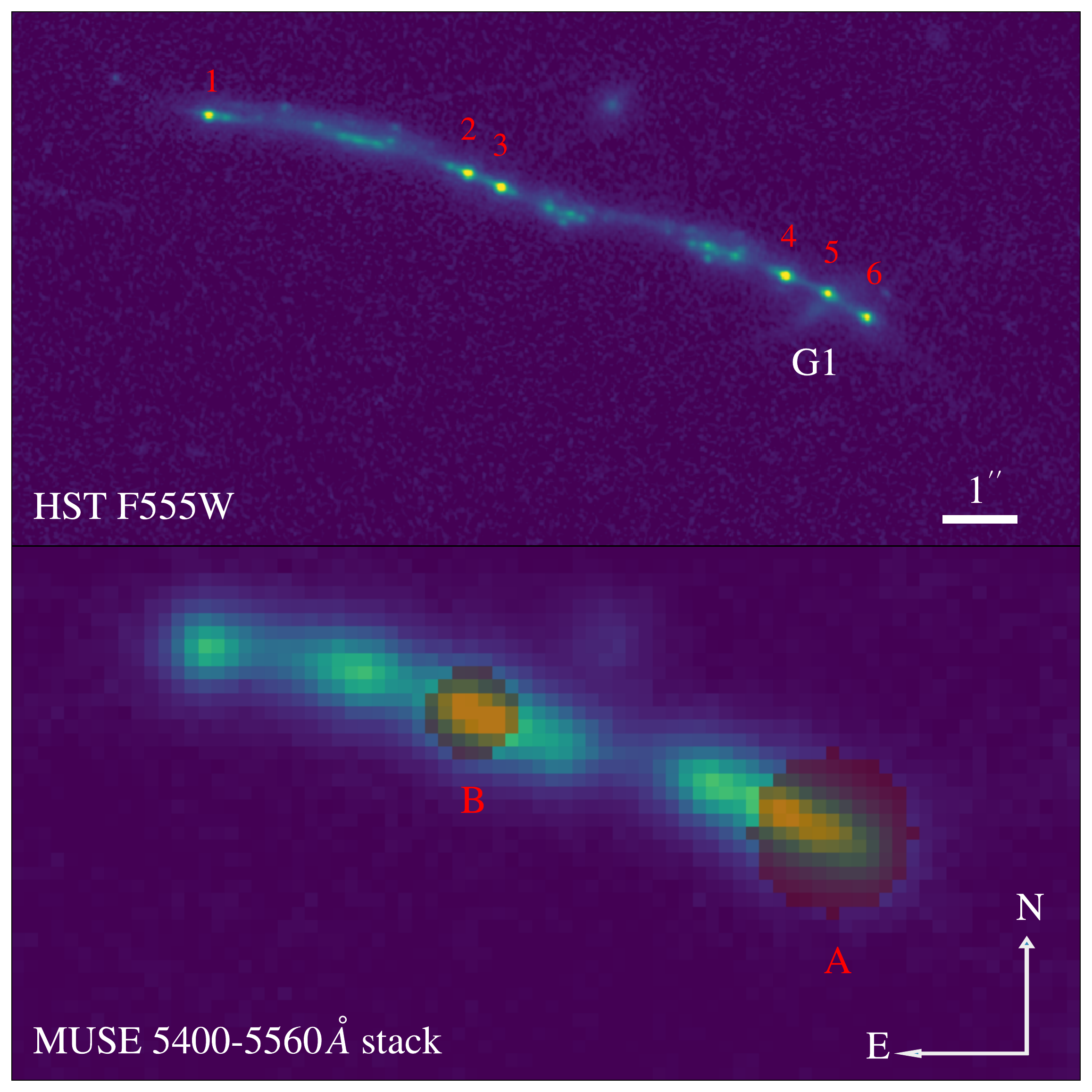}
    \caption{The same field of view through the HST F555W filter (top panel) and in the MUSE IFU image stacked across the wavelength range $5400$--$5560$\AA\,(bottom panel). The field of view shows Arc 1 of the Sunburst Arc, along which multiple lensed images of the LyC Knot denoted as Images 1 through 6 are clearly seen (\citep[we follow the denotation of][]{rivera2019gravitational}). The faint foreground galaxy G1 appears as a low-surface-brightness, elongated ellipsoid whose outskirt overlaps with the line of sight toward Image 5. In the MUSE IFU image, we show Apertures A and B which we use to extract the continuum SED of the galaxy G1.}
    \label{fig:arc}
\end{figure}

To infer its microlens population, we would like to model the stellar population of the foreground galaxy G1. To that end, we again choose to fit the stellar SED continuum using the MUSE IFU data, which offer richer information than from the photometry of wide HST filters. However, the seeing limited spatial resolution prevents us from directly separating the light of G1 from that of the nearby lensed images (Image 4, 5 and 6) of the LyC Knot, unlike in the case of wide filter HST images (\reffig{arc}).

To overcome this problem, we define two circular apertures and compute the integrated spectra enclosed within them. Aperture A has a radius of $1.2\arcsec$ and has its integrated flux dominated by Images 4, 5, and 6 of the LyC Knot, as well as the intervening G1. Aperture B has a radius of $0.38\arcsec$ and has its flux dominated by Images 2 and 3 of the LyC Knot. These aperture radii have been chosen to minimize contamination from nearby surface brightness features along the arc. We measure and subtract any uniform background, for Apertures A and B, respectively, by positioning the same aperture at a nearby empty place within the field of view.
Under the assumption that fluxes from the various lensed images of the LyC Knot have identical spectral shapes, we subtract a suitable rescaling of the Aperture B flux from Aperture A flux such that the residuals reflect the contribution from G1. 

\begin{figure*}
	\includegraphics[width=\textwidth]{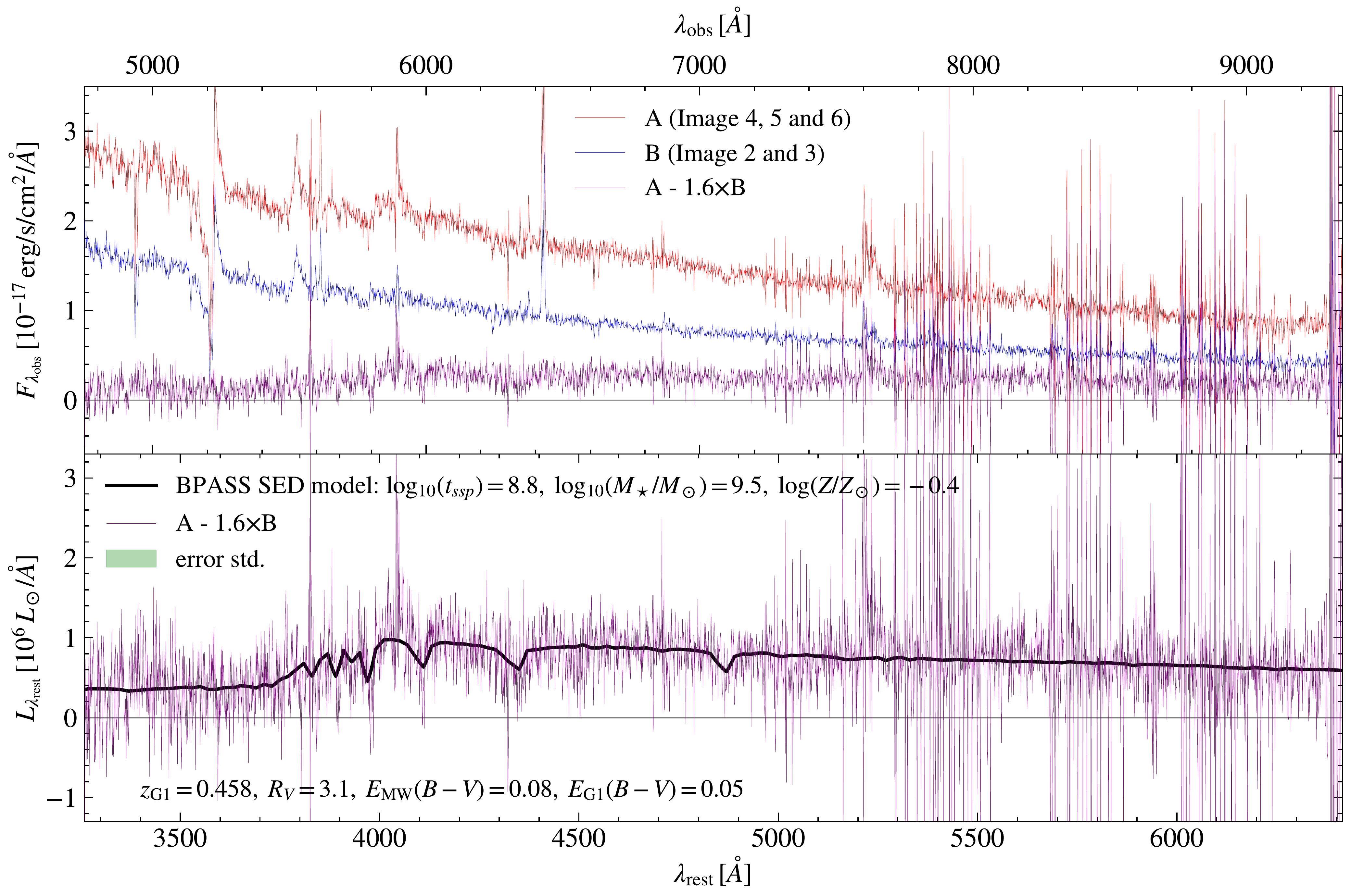}
    \caption{Modeling of the stellar population of the foreground galaxy G1 at $z_{\rm G1}=0.458$ using MUSE IFU data. {\it Top}: Observed-frame flux density integrated over Aperture A (red; containing Images 4, 5 and 6, and G1), Aperture B (blue; containing Images 2 and 3), and the subtraction of latter from the former (magenta). {\it Bottom}: Fitting of G1 rest-frame luminosity density (magenta; same as in top panel) using \textsc{BPASS} synthetic SEDs. We show the best-fit SSP (black) at an age $t_{\rm SSP} \approx 600\,$Myr, a metallicity $Z=0.4\,Z_\odot$, and a total stellar mass $M_\star=3\times 10^9\,\Msun$. Curves are dust reddened from within the Milky Way with $E_{\rm MW}(B-V)=0.08$~\citep{schlafly2011measuring}, and from within G1 with $E_{\rm G1}(B-V)=0.05$. The green shaded band indicates the size of the statistical uncertainty including sky background contamination. Many strong and sharp spectral features are due to instrumental artifacts or sky lines unrelated to the astrophysical sources.}
    \label{fig:G1_MUSE_SED_fit}
\end{figure*}

When 1.6 times of the Aperture B flux is subtracted from the Aperture A flux, as shown in the top panel of \reffig{G1_MUSE_SED_fit}, we completely remove the major spectral features of the LyC Knot, which include C IV $\lambda\lambda$1548,1550~\AA, He II $\lambda$1640~\AA, O III] $\lambda\lambda$1661,1666~\AA, and C III] $\lambda\lambda$1907,1909~\AA~\citep{vanzella2020ionizing}. The remaining flux, which supposedly comes from G1, exhibit Balmer absorption features H$\epsilon$, H$\delta$ and H$\gamma$, from which we infer a redshift $z_{\rm G1}=0.458$. This is close to, but slightly larger, than the spectroscopic redshift derived for the cluster lens PSZ1-G311 $z_l=0.443$~\citep{dahle2016discovery}, suggesting that G1 is not a cluster member galaxy. In any case, for the purpose of quantifying the intervening microlensing effects toward Image 5 under the influence of macro lensing by the galaxy cluster PSZ1-G311, it is a good approximation to assume that G1 has the same redshift as PSZ1-G311.

For a control test, we now center Aperture A at Image 1 and repeat the same procedure. This time, we remove the major spectral features by subtracting from its flux 0.47 times the flux of Aperture B. The residuals are more than a factor of 5 smaller than the magenta curve in \reffig{G1_MUSE_SED_fit} for G1. The small residuals are biased toward positive values probably because even away from the images of the LyC Knot the Sunburst Arc itself has a finite surface brightness, which is not expected to be removed by the subtraction method. This verifies that the continuum SED we have extracted for G1 is robust and is not severely contaminated by other sources.

Due to possible extra dust reddening by G1, Image 5 may differ from the other lensed images of the LyC Knot in the SED shape, which is unaccounted for in the above subtraction procedure. However, dust reddening due to G1 is likely mild $E_{\rm G1}(B-V) \simeq 0.05$ for $R_V=3.1$, according to the SED fitting of G1 as we will show next, and also because the line of sight toward Image 5 intersects only the outskirt of G1.

The subtraction residuals are dereddened assuming a Milky Way dust attenuation law $R_V=3.1$ and $E_{\rm MW}(B-V)=0.08$ along the line of sight according to the model of \cite{schlafly2011measuring}, redshifted to the source frame, and are then fitted to \textsc{BPASS}~\citep{eldridge2017binary, stanway2018re} synthetic SEDs for a simple stellar population (SSP). The \textsc{BPASS} synthetic SEDs are for a double power-law initial mass function (IMF)~\citep{2001MNRAS.322..231K} which has a slope $\alpha = -\rmd\log N/\rmd\log m = 1.30$ for main sequence masses $0.1\,\Msun < m < 0.5\,\Msun$, and a steepened slope $\alpha=2.35$ for $0.5\,\Msun < m < 100\,\Msun$. 

The SED continuum is best fit by an SSP of $t_{\rm SSP}=630\,$Myr old at sub-solar metallicity $Z = 0.4\,Z_\odot$, with some additional amount of local dust reddening $R_V=3.1$ and $E_{\rm G1}(B-V)=0.05$ (bottom panel of \reffig{G1_MUSE_SED_fit}). When the stellar population is allowed to comprise multiple components of different ages $10\,{\rm Myr} < t_{\rm SSP} < 10\,{\rm Gyr}$, and of different metallicities $0.05 < Z/Z_{\odot} < 2$, the fitting results remain stable and point toward the same SSP. When the source-frame dust reddening is varied $0.0 < E_{\rm G1}(B-V) < 0.2$, the best-fit metallicity alters, but the stellar mass and age only change by small amounts. The best-fit stellar population also reproduce the photometric magnitudes of G1 in the wide HST filters.

Due to blending, it is difficult to directly measure the surface brightness of G1 at the centroid of Image 5. Instead, we assume that the isophotes of G1 have ellipsoidal symmetry, and find the line of sight whose location is symmetric about the G1 center (peak of surface brightness) with the one toward Image 5. We assume that the surface brightness along that line of sight equals that along the line of sight toward Image 5.

As listed in \reftab{kappa_star}, several SSP solutions that provide a good fit all predict an average contribution to the lensing convergence from G1 member stars toward Image 5, $\kappa_\star = \Sigma_\star/\Sigma_{\rm crit} \approx 1$--$3\%$. Here $\Sigma_\star$ is the stellar surface mass density, and $\Sigma_{\rm crit}=(c^2/4\,\pi\,G)\,(D_S/D_L\,D_{LS})=2.0\times 10^9\,\Msun/{\rm kpc}^2$ is the critical surface density, with $D_L$, $D_S$ and $D_{LS}$ being the angular diameter distances to the lens plane at $z_l=0.44$, to the source plane at $z_s = 2.37$, and from the lens plane to the source plane. Variation in metallicity, stellar age and local dust reddening in these SSP solutions only leads to moderate changes in $\kappa_\star$, while the most noticeable uncertainty comes from the choice of HST filter used to estimate the surface brightness. This may be due to the filter dependence of the point spread function (PSF). 

The numbers quoted in \reftab{kappa_star} are not corrected for stellar mass loss. According to stellar population synthesis codes (e.g. Flexible Stellar Population Synthesis (FSPS)~\citep{conroy2009propagation, conroy2010propagation}), at an age $\log_{10}(t_{\rm SSP}/{\rm yr})=8.7$--$8.8$ about $70\%$ of the total initial stellar mass (including stellar remnants) survives either mass loss or stellar explosion. Applying this correction, we estimate that $\kappa_\star=0.7$--$2\%$.

\begin{table*}
	\centering
	\caption{Several comparably good fits to the stellar population of G1 using the \textsc{BPASS} models. We list the corresponding predictions for the stellar contribution to the lensing convergence $\kappa_\star$ along the line of sight to Image 5 as inferred from the HST images in the F555W, F606W and F814W filters, respectively. Multiple visits of the same filter are averaged over. A Milky Way like dust extinction law $R_V =3.1$ is assumed when applying source-frame dust reddening set by $E_{\rm G1}(B-V)$. Mass loss effects from stellar evolution are not accounted for in the table, which we estimate cause a further $70\%$ reduction in $\kappa_\star$.}
	\label{tab:kappa_star}
	\begin{tabular}{cccccc} 
		\hline
		\hline
		$\log_{10}(Z/Z_\odot)$ & $\log_{10}(t_{\rm SSP}/{\rm yr})$ & $E_{\rm G1}(B-V)$ & $\kappa_\star$ (from F555W) & $\kappa_\star$ (from F606W) & $\kappa_\star$ (from F814W) \\
		\hline
		-0.30 & 8.8 & 0.00 & 0.012 & 0.025 & 0.016 \\
		-0.30 & 8.8 & 0.02 & 0.013 & 0.027 & 0.017 \\
		-0.40 & 8.8 & 0.05 & 0.013 & 0.029 & 0.018 \\
		-0.15 & 8.7 & 0.08 & 0.014 & 0.029 & 0.018 \\
		-0.15 & 8.7 & 0.10 & 0.015 & 0.032 & 0.019 \\
		\hline
	\end{tabular}
\end{table*}

\bibliographystyle{mnras}
\bibliography{refs_new.bib}

\begin{thebibliography}{}
\makeatletter
\relax
\def\mn@urlcharsother{\let\do\@makeother \do\$\do\&\do\#\do\^\do\_\do\%\do\~}
\def\mn@doi{\begingroup\mn@urlcharsother \@ifnextchar [ {\mn@doi@}
  {\mn@doi@[]}}
\def\mn@doi@[#1]#2{\def\@tempa{#1}\ifx\@tempa\@empty \href
  {http://dx.doi.org/#2} {doi:#2}\else \href {http://dx.doi.org/#2} {#1}\fi
  \endgroup}
\def\mn@eprint#1#2{\mn@eprint@#1:#2::\@nil}
\def\mn@eprint@arXiv#1{\href {http://arxiv.org/abs/#1} {{\tt arXiv:#1}}}
\def\mn@eprint@dblp#1{\href {http://dblp.uni-trier.de/rec/bibtex/#1.xml}
  {dblp:#1}}
\def\mn@eprint@#1:#2:#3:#4\@nil{\def\@tempa {#1}\def\@tempb {#2}\def\@tempc
  {#3}\ifx \@tempc \@empty \let \@tempc \@tempb \let \@tempb \@tempa \fi \ifx
  \@tempb \@empty \def\@tempb {arXiv}\fi \@ifundefined
  {mn@eprint@\@tempb}{\@tempb:\@tempc}{\expandafter \expandafter \csname
  mn@eprint@\@tempb\endcsname \expandafter{\@tempc}}}

\bibitem[\protect\citeauthoryear{{Alcock} et~al.,}{{Alcock}
  et~al.}{1999}]{AlcockEtAl1999DiffImage}
{Alcock} C.,  et~al., 1999, \mn@doi [\apj] {10.1086/307567}, \href
  {https://ui.adsabs.harvard.edu/abs/1999ApJ...521..602A} {521, 602}

\bibitem[\protect\citeauthoryear{{Alcock} et~al.,}{{Alcock}
  et~al.}{2001}]{alcock2001macho}
{Alcock} C.,  et~al., 2001, \mn@doi [\apjl] {10.1086/319636}, \href
  {https://ui.adsabs.harvard.edu/abs/2001ApJ...550L.169A} {550, L169}

\bibitem[\protect\citeauthoryear{{Ali-Ha{\"\i}moud} \&
  {Kamionkowski}}{{Ali-Ha{\"\i}moud} \&
  {Kamionkowski}}{2017}]{AliHaimoud2017machoCMB}
{Ali-Ha{\"\i}moud} Y.,  {Kamionkowski} M.,  2017, \mn@doi [\prd]
  {10.1103/PhysRevD.95.043534}, \href
  {https://ui.adsabs.harvard.edu/abs/2017PhRvD..95d3534A} {95, 043534}

\bibitem[\protect\citeauthoryear{{Bird}, {Cholis}, {Mu{\~n}oz},
  {Ali-Ha{\"\i}moud}, {Kamionkowski}, {Kovetz}, {Raccanelli}  \&
  {Riess}}{{Bird} et~al.}{2016}]{bird2016did}
{Bird} S.,  {Cholis} I.,  {Mu{\~n}oz} J.~B.,  {Ali-Ha{\"\i}moud} Y.,
  {Kamionkowski} M.,  {Kovetz} E.~D.,  {Raccanelli} A.,   {Riess} A.~G.,  2016,
  \mn@doi [\prl] {10.1103/PhysRevLett.116.201301}, \href
  {https://ui.adsabs.harvard.edu/abs/2016PhRvL.116t1301B} {116, 201301}

\bibitem[\protect\citeauthoryear{{Blandford} \& {Narayan}}{{Blandford} \&
  {Narayan}}{1986}]{blandford1986fermat}
{Blandford} R.,  {Narayan} R.,  1986, \mn@doi [\apj] {10.1086/164709}, \href
  {https://ui.adsabs.harvard.edu/abs/1986ApJ...310..568B} {310, 568}

\bibitem[\protect\citeauthoryear{{Brandt}}{{Brandt}}{2016}]{Brandt2016MACHOconstraint}
{Brandt} T.~D.,  2016, \mn@doi [\apjl] {10.3847/2041-8205/824/2/L31}, \href
  {https://ui.adsabs.harvard.edu/abs/2016ApJ...824L..31B} {824, L31}

\bibitem[\protect\citeauthoryear{{Carr} \& {Hawking}}{{Carr} \&
  {Hawking}}{1974}]{carr1974black}
{Carr} B.~J.,  {Hawking} S.~W.,  1974, \mn@doi [\mnras]
  {10.1093/mnras/168.2.399}, \href
  {https://ui.adsabs.harvard.edu/abs/1974MNRAS.168..399C} {168, 399}

\bibitem[\protect\citeauthoryear{{Chang} \& {Refsdal}}{{Chang} \&
  {Refsdal}}{1984}]{chang1984star}
{Chang} K.,  {Refsdal} S.,  1984, \aap, \href
  {https://ui.adsabs.harvard.edu/abs/1984A&A...132..168C} {132, 168}

\bibitem[\protect\citeauthoryear{{Chapline}}{{Chapline}}{1975}]{chapline1975cosmological}
{Chapline} G.~F.,  1975, \mn@doi [\nat] {10.1038/253251a0}, \href
  {https://ui.adsabs.harvard.edu/abs/1975Natur.253..251C} {253, 251}

\bibitem[\protect\citeauthoryear{{Chen} et~al.,}{{Chen}
  et~al.}{2019}]{Chen:2019ncy}
{Chen} W.,  et~al., 2019, \mn@doi [\apj] {10.3847/1538-4357/ab297d}, \href
  {https://ui.adsabs.harvard.edu/abs/2019ApJ...881....8C} {881, 8}

\bibitem[\protect\citeauthoryear{{Chisholm}, {Rigby}, {Bayliss}, {Berg},
  {Dahle}, {Gladders}  \& {Sharon}}{{Chisholm}
  et~al.}{2019}]{2019ApJ...882..182C}
{Chisholm} J.,  {Rigby} J.~R.,  {Bayliss} M.,  {Berg} D.~A.,  {Dahle} H.,
  {Gladders} M.,   {Sharon} K.,  2019, \mn@doi [\apj]
  {10.3847/1538-4357/ab3104}, \href
  {https://ui.adsabs.harvard.edu/abs/2019ApJ...882..182C} {882, 182}

\bibitem[\protect\citeauthoryear{{Conroy} \& {Gunn}}{{Conroy} \&
  {Gunn}}{2010}]{conroy2010propagation}
{Conroy} C.,  {Gunn} J.~E.,  2010, \mn@doi [\apj]
  {10.1088/0004-637X/712/2/833}, \href
  {https://ui.adsabs.harvard.edu/abs/2010ApJ...712..833C} {712, 833}

\bibitem[\protect\citeauthoryear{{Conroy}, {Gunn}  \& {White}}{{Conroy}
  et~al.}{2009}]{conroy2009propagation}
{Conroy} C.,  {Gunn} J.~E.,   {White} M.,  2009, \mn@doi [\apj]
  {10.1088/0004-637X/699/1/486}, \href
  {https://ui.adsabs.harvard.edu/abs/2009ApJ...699..486C} {699, 486}

\bibitem[\protect\citeauthoryear{{Conroy} et~al.,}{{Conroy}
  et~al.}{2018}]{Controy2018M51variability}
{Conroy} C.,  et~al., 2018, \mn@doi [\apj] {10.3847/1538-4357/aad460}, \href
  {https://ui.adsabs.harvard.edu/abs/2018ApJ...864..111C} {864, 111}

\bibitem[\protect\citeauthoryear{{Crotts}}{{Crotts}}{1992}]{Crotts1992M31}
{Crotts} A. P.~S.,  1992, \mn@doi [\apjl] {10.1086/186602}, \href
  {https://ui.adsabs.harvard.edu/abs/1992ApJ...399L..43C} {399, L43}

\bibitem[\protect\citeauthoryear{{Crotts} \& {Tomaney}}{{Crotts} \&
  {Tomaney}}{1996}]{CrottsTomaney1996M31}
{Crotts} A. P.~S.,  {Tomaney} A.~B.,  1996, \mn@doi [\apjl] {10.1086/310405},
  \href {https://ui.adsabs.harvard.edu/abs/1996ApJ...473L..87C} {473, L87}

\bibitem[\protect\citeauthoryear{{Crowther} et~al.,}{{Crowther}
  et~al.}{2016}]{crowther2016r136}
{Crowther} P.~A.,  et~al., 2016, \mn@doi [\mnras] {10.1093/mnras/stw273}, \href
  {https://ui.adsabs.harvard.edu/abs/2016MNRAS.458..624C} {458, 624}

\bibitem[\protect\citeauthoryear{{Dahle} et~al.,}{{Dahle}
  et~al.}{2016}]{dahle2016discovery}
{Dahle} H.,  et~al., 2016, \mn@doi [\aap] {10.1051/0004-6361/201628297}, \href
  {https://ui.adsabs.harvard.edu/abs/2016A&A...590L...4D} {590, L4}

\bibitem[\protect\citeauthoryear{{Dai}, {Venumadhav}, {Kaurov}  \&
  {Miralda-Escud}}{{Dai} et~al.}{2018}]{2018ApJ...867...24D}
{Dai} L.,  {Venumadhav} T.,  {Kaurov} A.~A.,   {Miralda-Escud} J.,  2018,
  \mn@doi [\apj] {10.3847/1538-4357/aae478}, \href
  {http://adsabs.harvard.edu/abs/2018ApJ...867...24D} {867, 24}

\bibitem[\protect\citeauthoryear{{Dai} et~al.,}{{Dai}
  et~al.}{2020}]{Dai2020S1226}
{Dai} L.,  et~al., 2020, \mn@doi [\mnras] {10.1093/mnras/staa1355}, \href
  {https://ui.adsabs.harvard.edu/abs/2020MNRAS.495.3192D} {495, 3192}

\bibitem[\protect\citeauthoryear{{Deguchi} \& {Watson}}{{Deguchi} \&
  {Watson}}{1987}]{PhysRevLett.59.2814}
{Deguchi} S.,  {Watson} W.~D.,  1987, \mn@doi [\prl]
  {10.1103/PhysRevLett.59.2814}, \href
  {https://ui.adsabs.harvard.edu/abs/1987PhRvL..59.2814D} {59, 2814}

\bibitem[\protect\citeauthoryear{{Diego}}{{Diego}}{2019}]{Diego2019HighMuUniverse}
{Diego} J.~M.,  2019, \mn@doi [\aap] {10.1051/0004-6361/201833670}, \href
  {https://ui.adsabs.harvard.edu/abs/2019A&A...625A..84D} {625, A84}

\bibitem[\protect\citeauthoryear{{Diego} et~al.,}{{Diego}
  et~al.}{2018}]{2018ApJ...857...25D}
{Diego} J.~M.,  et~al., 2018, \mn@doi [\apj] {10.3847/1538-4357/aab617}, \href
  {http://adsabs.harvard.edu/abs/2018ApJ...857...25D} {857, 25}

\bibitem[\protect\citeauthoryear{{Dunstall} et~al.,}{{Dunstall}
  et~al.}{2015}]{Dunstall2015MultipleBstars}
{Dunstall} P.~R.,  et~al., 2015, \mn@doi [\aap] {10.1051/0004-6361/201526192},
  \href {https://ui.adsabs.harvard.edu/abs/2015A&A...580A..93D} {580, A93}

\bibitem[\protect\citeauthoryear{{Eldridge}, {Stanway}, {Xiao}, {McClelland },
  {Taylor}, {Ng}, {Greis}  \& {Bray}}{{Eldridge}
  et~al.}{2017}]{eldridge2017binary}
{Eldridge} J.~J.,  {Stanway} E.~R.,  {Xiao} L.,  {McClelland } L.~A.~S.,
  {Taylor} G.,  {Ng} M.,  {Greis} S.~M.~L.,   {Bray} J.~C.,  2017, \mn@doi
  [\pasa] {10.1017/pasa.2017.51}, \href
  {https://ui.adsabs.harvard.edu/abs/2017PASA...34...58E} {34, e058}

\bibitem[\protect\citeauthoryear{{Fleury} \& {Garc{\'\i}a-Bellido}}{{Fleury} \&
  {Garc{\'\i}a-Bellido}}{2020}]{fleury2020simple}
{Fleury} P.,  {Garc{\'\i}a-Bellido} J.,  2020, \mn@doi [Physics of the Dark
  Universe] {10.1016/j.dark.2020.100567}, \href
  {https://ui.adsabs.harvard.edu/abs/2020PDU....2900567F} {29, 100567}

\bibitem[\protect\citeauthoryear{{Gaburov}, {Gualandris}  \& {Portegies
  Zwart}}{{Gaburov} et~al.}{2008}]{Gaburov2008collI}
{Gaburov} E.,  {Gualandris} A.,   {Portegies Zwart} S.,  2008, \mn@doi [\mnras]
  {10.1111/j.1365-2966.2007.12731.x}, \href
  {https://ui.adsabs.harvard.edu/abs/2008MNRAS.384..376G} {384, 376}

\bibitem[\protect\citeauthoryear{{Gaburov}, {Lombardi}  \& {Portegies
  Zwart}}{{Gaburov} et~al.}{2010}]{Gaburov2010collII}
{Gaburov} E.,  {Lombardi} James~C. J.,   {Portegies Zwart} S.,  2010, \mn@doi
  [\mnras] {10.1111/j.1365-2966.2009.15900.x}, \href
  {https://ui.adsabs.harvard.edu/abs/2010MNRAS.402..105G} {402, 105}

\bibitem[\protect\citeauthoryear{{Gil-Merino} \& {Lewis}}{{Gil-Merino} \&
  {Lewis}}{2006}]{GilMerinoLewis2006StarFormingRegion}
{Gil-Merino} R.,  {Lewis} G.~F.,  2006, \mn@doi [\apj] {10.1086/502797}, \href
  {https://ui.adsabs.harvard.edu/abs/2006ApJ...643..260G} {643, 260}

\bibitem[\protect\citeauthoryear{{Goldstein}, {Nugent}, {Kasen}  \&
  {Collett}}{{Goldstein} et~al.}{2018}]{goldstein2018precise}
{Goldstein} D.~A.,  {Nugent} P.~E.,  {Kasen} D.~N.,   {Collett} T.~E.,  2018,
  \mn@doi [\apj] {10.3847/1538-4357/aaa975}, \href
  {https://ui.adsabs.harvard.edu/abs/2018ApJ...855...22G} {855, 22}

\bibitem[\protect\citeauthoryear{{Gott}}{{Gott}}{1981}]{GottIII1981QSOMicrolensing}
{Gott} J.~R. I.,  1981, \mn@doi [\apj] {10.1086/158576}, \href
  {https://ui.adsabs.harvard.edu/abs/1981ApJ...243..140G} {243, 140}

\bibitem[\protect\citeauthoryear{{Griest}}{{Griest}}{1991}]{griest1991galactic}
{Griest} K.,  1991, \mn@doi [\apj] {10.1086/169575}, \href
  {https://ui.adsabs.harvard.edu/abs/1991ApJ...366..412G} {366, 412}

\bibitem[\protect\citeauthoryear{{Griest}, {Cieplak}  \& {Lehner}}{{Griest}
  et~al.}{2013}]{griest2013new}
{Griest} K.,  {Cieplak} A.~M.,   {Lehner} M.~J.,  2013, \mn@doi [\prl]
  {10.1103/PhysRevLett.111.181302}, \href
  {https://ui.adsabs.harvard.edu/abs/2013PhRvL.111r1302G} {111, 181302}

\bibitem[\protect\citeauthoryear{{Je{\v{r}}{\'a}bkov{\'a}}, {Kroupa},
  {Dabringhausen}, {Hilker}  \& {Bekki}}{{Je{\v{r}}{\'a}bkov{\'a}}
  et~al.}{2017}]{jevrabkova2017formation}
{Je{\v{r}}{\'a}bkov{\'a}} T.,  {Kroupa} P.,  {Dabringhausen} J.,  {Hilker} M.,
   {Bekki} K.,  2017, \mn@doi [\aap] {10.1051/0004-6361/201731240}, \href
  {https://ui.adsabs.harvard.edu/abs/2017A&A...608A..53J} {608, A53}

\bibitem[\protect\citeauthoryear{{Ji}, {Kovetz}  \& {Kamionkowski}}{{Ji}
  et~al.}{2018}]{JiKovetzKamionkowski2018GRBMicrolensing}
{Ji} L.,  {Kovetz} E.~D.,   {Kamionkowski} M.,  2018, \mn@doi [\prd]
  {10.1103/PhysRevD.98.123523}, \href
  {https://ui.adsabs.harvard.edu/abs/2018PhRvD..98l3523J} {98, 123523}

\bibitem[\protect\citeauthoryear{{Jung} \& {Shin}}{{Jung} \&
  {Shin}}{2019}]{PhysRevLett.122.041103}
{Jung} S.,  {Shin} C.~S.,  2019, \mn@doi [\prl]
  {10.1103/PhysRevLett.122.041103}, \href
  {https://ui.adsabs.harvard.edu/abs/2019PhRvL.122d1103J} {122, 041103}

\bibitem[\protect\citeauthoryear{{Katz}, {Balbus}  \& {Paczynski}}{{Katz}
  et~al.}{1986}]{1986ApJ...306....2K}
{Katz} N.,  {Balbus} S.,   {Paczynski} B.,  1986, \mn@doi [\apj]
  {10.1086/164313}, \href
  {https://ui.adsabs.harvard.edu/abs/1986ApJ...306....2K} {306, 2}

\bibitem[\protect\citeauthoryear{{Kaurov}, {Dai}, {Venumadhav},
  {Miralda-Escud{\'e}}  \& {Frye}}{{Kaurov} et~al.}{2019}]{2019ApJ...880...58K}
{Kaurov} A.~A.,  {Dai} L.,  {Venumadhav} T.,  {Miralda-Escud{\'e}} J.,   {Frye}
  B.,  2019, \mn@doi [\apj] {10.3847/1538-4357/ab2888}, \href
  {https://ui.adsabs.harvard.edu/abs/2019ApJ...880...58K} {880, 58}

\bibitem[\protect\citeauthoryear{{Kelly} et~al.,}{{Kelly}
  et~al.}{2018}]{2018NatAs...2..334K}
{Kelly} P.~L.,  et~al., 2018, \mn@doi [Nature Astronomy]
  {10.1038/s41550-018-0430-3}, \href
  {http://adsabs.harvard.edu/abs/2018NatAs...2..334K} {2, 334}

\bibitem[\protect\citeauthoryear{{Kelly} et~al.,}{{Kelly}
  et~al.}{2019}]{Kelly2019hstflashlight}
{Kelly} P.,  et~al., 2019, {Flashlights: Many Extremely Magnified Individual
  Stars as Probes of Dark Matter and Stellar Populations to Redshift z 2}, HST
  Proposal

\bibitem[\protect\citeauthoryear{{Kofman}, {Kaiser}, {Lee}  \&
  {Babul}}{{Kofman} et~al.}{1997}]{KofmanEtAl1997TwoD}
{Kofman} L.,  {Kaiser} N.,  {Lee} M.~H.,   {Babul} A.,  1997, \mn@doi [\apj]
  {10.1086/304791}, \href
  {https://ui.adsabs.harvard.edu/abs/1997ApJ...489..508K} {489, 508}

\bibitem[\protect\citeauthoryear{{Kroupa}}{{Kroupa}}{2001a}]{kroupa2001variation}
{Kroupa} P.,  2001a, \mn@doi [\mnras] {10.1046/j.1365-8711.2001.04022.x}, \href
  {https://ui.adsabs.harvard.edu/abs/2001MNRAS.322..231K} {322, 231}

\bibitem[\protect\citeauthoryear{{Kroupa}}{{Kroupa}}{2001b}]{2001MNRAS.322..231K}
{Kroupa} P.,  2001b, \mn@doi [\mnras] {10.1046/j.1365-8711.2001.04022.x}, \href
  {https://ui.adsabs.harvard.edu/abs/2001MNRAS.322..231K} {322, 231}

\bibitem[\protect\citeauthoryear{{Laur}, {Kolka}, {Eenm{\"a}e}, {Tuvikene}  \&
  {Leedj{\"a}rv}}{{Laur} et~al.}{2017}]{Laur2017variability}
{Laur} J.,  {Kolka} I.,  {Eenm{\"a}e} T.,  {Tuvikene} T.,   {Leedj{\"a}rv} L.,
  2017, \mn@doi [\aap] {10.1051/0004-6361/201629395}, \href
  {https://ui.adsabs.harvard.edu/abs/2017A&A...598A.108L} {598, A108}

\bibitem[\protect\citeauthoryear{{Lee}, {Babul}, {Kofman}  \& {Kaiser}}{{Lee}
  et~al.}{1997}]{LeeEtAl1997ThreeD}
{Lee} M.~H.,  {Babul} A.,  {Kofman} L.,   {Kaiser} N.,  1997, \mn@doi [\apj]
  {10.1086/304792}, \href
  {https://ui.adsabs.harvard.edu/abs/1997ApJ...489..522L} {489, 522}

\bibitem[\protect\citeauthoryear{{Lef{\`e}vre}, {Marchenko}, {Moffat}  \&
  {Acker}}{{Lef{\`e}vre} et~al.}{2009}]{Lefevre2009OBstarsHIPPARCOS}
{Lef{\`e}vre} L.,  {Marchenko} S.~V.,  {Moffat} A.~F.~J.,   {Acker} A.,  2009,
  \mn@doi [\aap] {10.1051/0004-6361/200912304}, \href
  {https://ui.adsabs.harvard.edu/abs/2009A&A...507.1141L} {507, 1141}

\bibitem[\protect\citeauthoryear{{Lewis} \& {Ibata}}{{Lewis} \&
  {Ibata}}{2001}]{LewisIbata2000MACHO}
{Lewis} G.~F.,  {Ibata} R.~A.,  2001, \mn@doi [\apj] {10.1086/319041}, \href
  {https://ui.adsabs.harvard.edu/abs/2001ApJ...549...46L} {549, 46}

\bibitem[\protect\citeauthoryear{{Lewis} \& {Irwin}}{{Lewis} \&
  {Irwin}}{1995}]{GeraintIrwin1995AmpPDF}
{Lewis} G.~F.,  {Irwin} M.~J.,  1995, \mn@doi [\mnras]
  {10.1093/mnras/276.1.103}, \href
  {https://ui.adsabs.harvard.edu/abs/1995MNRAS.276..103L} {276, 103}

\bibitem[\protect\citeauthoryear{{Lewis} \& {Irwin}}{{Lewis} \&
  {Irwin}}{1996}]{GeraintIrwin1995TemporalAnalysis}
{Lewis} G.~F.,  {Irwin} M.~J.,  1996, \mn@doi [\mnras]
  {10.1093/mnras/283.1.225}, \href
  {https://ui.adsabs.harvard.edu/abs/1996MNRAS.283..225L} {283, 225}

\bibitem[\protect\citeauthoryear{{Lewis}, {Miralda-Escude}, {Richardson}  \&
  {Wambsganss}}{{Lewis} et~al.}{1993}]{lewis1993microlensing}
{Lewis} G.~F.,  {Miralda-Escude} J.,  {Richardson} D.~C.,   {Wambsganss} J.,
  1993, \mn@doi [\mnras] {10.1093/mnras/261.3.647}, \href
  {https://ui.adsabs.harvard.edu/abs/1993MNRAS.261..647L} {261, 647}

\bibitem[\protect\citeauthoryear{{Lewis}, {Ibata}  \& {Wyithe}}{{Lewis}
  et~al.}{2000}]{LewisIbataWyithe2000MACHO}
{Lewis} G.~F.,  {Ibata} R.~A.,   {Wyithe} J. S.~B.,  2000, \mn@doi [\apjl]
  {10.1086/312916}, \href
  {https://ui.adsabs.harvard.edu/abs/2000ApJ...542L...9L} {542, L9}

\bibitem[\protect\citeauthoryear{{Liao}, {Zhang}, {Li}  \& {Gao}}{{Liao}
  et~al.}{2020}]{2020arXiv200313349L}
{Liao} K.,  {Zhang} S.~B.,  {Li} Z.,   {Gao} H.,  2020, \mn@doi [\apjl]
  {10.3847/2041-8213/ab963e}, \href
  {https://ui.adsabs.harvard.edu/abs/2020ApJ...896L..11L} {896, L11}

\bibitem[\protect\citeauthoryear{{Lin} \& {Mohr}}{{Lin} \&
  {Mohr}}{2004}]{lin2004k}
{Lin} Y.-T.,  {Mohr} J.~J.,  2004, \mn@doi [\apj] {10.1086/425412}, \href
  {https://ui.adsabs.harvard.edu/abs/2004ApJ...617..879L} {617, 879}

\bibitem[\protect\citeauthoryear{{Ma}, {Hopkins}, {Kasen}, {Quataert},
  {Faucher-Gigu{\`e}re}, {Kere{\v{s}}}, {Murray}  \& {Strom}}{{Ma}
  et~al.}{2016}]{Ma2016BinaryReionization}
{Ma} X.,  {Hopkins} P.~F.,  {Kasen} D.,  {Quataert} E.,  {Faucher-Gigu{\`e}re}
  C.-A.,  {Kere{\v{s}}} D.,  {Murray} N.,   {Strom} A.,  2016, \mn@doi [\mnras]
  {10.1093/mnras/stw941}, \href
  {https://ui.adsabs.harvard.edu/abs/2016MNRAS.459.3614M} {459, 3614}

\bibitem[\protect\citeauthoryear{{Marchandon} \& {Nottale}}{{Marchandon} \&
  {Nottale}}{1991}]{1991A&A...251..393M}
{Marchandon} S.,  {Nottale} L.,  1991, \aap, \href
  {https://ui.adsabs.harvard.edu/abs/1991A&A...251..393M} {251, 393}

\bibitem[\protect\citeauthoryear{{Mediavilla}, {Jim{\'e}nez-Vicente},
  {Mu{\~n}oz}, {Vives-Arias}  \& {Calder{\'o}n-Infante}}{{Mediavilla}
  et~al.}{2017}]{mediavilla2017limits}
{Mediavilla} E.,  {Jim{\'e}nez-Vicente} J.,  {Mu{\~n}oz} J.~A.,  {Vives-Arias}
  H.,   {Calder{\'o}n-Infante} J.,  2017, \mn@doi [\apjl]
  {10.3847/2041-8213/aa5dab}, \href
  {https://ui.adsabs.harvard.edu/abs/2017ApJ...836L..18M} {836, L18}

\bibitem[\protect\citeauthoryear{{Meszaros}}{{Meszaros}}{1974}]{meszaros1974behaviour}
{Meszaros} P.,  1974, \aap, \href
  {https://ui.adsabs.harvard.edu/abs/1974A&A....37..225M} {37, 225}

\bibitem[\protect\citeauthoryear{{Miralda-Escud\'e}}{{Miralda-Escud\'e}}{1991}]{1991ApJ...379...94M}
{Miralda-Escud\'e} J.,  1991, \mn@doi [\apj] {10.1086/170486}, \href
  {http://adsabs.harvard.edu/abs/1991ApJ...379...94M} {379, 94}

\bibitem[\protect\citeauthoryear{{Mu{\~n}oz}, {Kovetz}, {Dai}  \&
  {Kamionkowski}}{{Mu{\~n}oz} et~al.}{2016}]{Munoz:2016tmg}
{Mu{\~n}oz} J.~B.,  {Kovetz} E.~D.,  {Dai} L.,   {Kamionkowski} M.,  2016,
  \mn@doi [\prl] {10.1103/PhysRevLett.117.091301}, \href
  {https://ui.adsabs.harvard.edu/abs/2016PhRvL.117i1301M} {117, 091301}

\bibitem[\protect\citeauthoryear{{Nakamura}, {Sasaki}, {Tanaka}  \&
  {Thorne}}{{Nakamura} et~al.}{1997}]{nakamura1997gravitational}
{Nakamura} T.,  {Sasaki} M.,  {Tanaka} T.,   {Thorne} K.~S.,  1997, \mn@doi
  [\apjl] {10.1086/310886}, \href
  {https://ui.adsabs.harvard.edu/abs/1997ApJ...487L.139N} {487, L139}

\bibitem[\protect\citeauthoryear{{Navarro}, {Frenk}  \& {White}}{{Navarro}
  et~al.}{1996}]{1996ApJ...462..563N}
{Navarro} J.~F.,  {Frenk} C.~S.,   {White} S.~D.~M.,  1996, \mn@doi [\apj]
  {10.1086/177173}, \href {http://adsabs.harvard.edu/abs/1996ApJ...462..563N}
  {462, 563}

\bibitem[\protect\citeauthoryear{{Navarro}, {Frenk}  \& {White}}{{Navarro}
  et~al.}{1997}]{navarro1997universal}
{Navarro} J.~F.,  {Frenk} C.~S.,   {White} S. D.~M.,  1997, \mn@doi [\apj]
  {10.1086/304888}, \href
  {https://ui.adsabs.harvard.edu/abs/1997ApJ...490..493N} {490, 493}

\bibitem[\protect\citeauthoryear{{Neindorf}}{{Neindorf}}{2003}]{Neindorf2003MicrolensingAutoCorr}
{Neindorf} B.,  2003, \mn@doi [\aap] {10.1051/0004-6361:20030098}, \href
  {https://ui.adsabs.harvard.edu/abs/2003A&A...404...83N} {404, 83}

\bibitem[\protect\citeauthoryear{{Niikura} et~al.,}{{Niikura}
  et~al.}{2019}]{2019NatAs...3..524N}
{Niikura} H.,  et~al., 2019, \mn@doi [Nature Astronomy]
  {10.1038/s41550-019-0723-1}, \href
  {https://ui.adsabs.harvard.edu/abs/2019NatAs...3..524N} {3, 524}

\bibitem[\protect\citeauthoryear{{Oguri} \& {Blandford}}{{Oguri} \&
  {Blandford}}{2009}]{OguriBlandford2009LargestEinsteinRadii}
{Oguri} M.,  {Blandford} R.~D.,  2009, \mn@doi [\mnras]
  {10.1111/j.1365-2966.2008.14154.x}, \href
  {https://ui.adsabs.harvard.edu/abs/2009MNRAS.392..930O} {392, 930}

\bibitem[\protect\citeauthoryear{{Oguri}, {Diego}, {Kaiser}, {Kelly}  \&
  {Broadhurst}}{{Oguri} et~al.}{2018}]{Oguri:2017ock}
{Oguri} M.,  {Diego} J.~M.,  {Kaiser} N.,  {Kelly} P.~L.,   {Broadhurst} T.,
  2018, \mn@doi [\prd] {10.1103/PhysRevD.97.023518}, \href
  {https://ui.adsabs.harvard.edu/abs/2018PhRvD..97b3518O} {97, 023518}

\bibitem[\protect\citeauthoryear{{Paczynski}}{{Paczynski}}{1986}]{paczynski1986gravitational}
{Paczynski} B.,  1986, \mn@doi [\apj] {10.1086/164140}, \href
  {https://ui.adsabs.harvard.edu/abs/1986ApJ...304....1P} {304, 1}

\bibitem[\protect\citeauthoryear{{Peacock}}{{Peacock}}{1982}]{peacock1982gravitational}
{Peacock} J.~A.,  1982, \mn@doi [\mnras] {10.1093/mnras/199.4.987}, \href
  {https://ui.adsabs.harvard.edu/abs/1982MNRAS.199..987P} {199, 987}

\bibitem[\protect\citeauthoryear{{Planck Collaboration} et~al.,}{{Planck
  Collaboration} et~al.}{2014}]{2014A&A...571A..29P}
{Planck Collaboration} et~al., 2014, \mn@doi [\aap]
  {10.1051/0004-6361/201321523}, \href
  {https://ui.adsabs.harvard.edu/abs/2014A&A...571A..29P} {571, A29}

\bibitem[\protect\citeauthoryear{{Priewe}, {Williams}, {Liesenborgs}, {Coe}  \&
  {Rodney}}{{Priewe} et~al.}{2017}]{Priewe2017LensModelMuError}
{Priewe} J.,  {Williams} L. L.~R.,  {Liesenborgs} J.,  {Coe} D.,   {Rodney}
  S.~A.,  2017, \mn@doi [\mnras] {10.1093/mnras/stw2785}, \href
  {https://ui.adsabs.harvard.edu/abs/2017MNRAS.465.1030P} {465, 1030}

\bibitem[\protect\citeauthoryear{{Rasia}, {Tormen}  \& {Moscardini}}{{Rasia}
  et~al.}{2004}]{rasia2004dynamical}
{Rasia} E.,  {Tormen} G.,   {Moscardini} L.,  2004, \mn@doi [\mnras]
  {10.1111/j.1365-2966.2004.07775.x}, \href
  {https://ui.adsabs.harvard.edu/abs/2004MNRAS.351..237R} {351, 237}

\bibitem[\protect\citeauthoryear{{Ricotti}}{{Ricotti}}{2002}]{Ricotti2002ClusterReionization}
{Ricotti} M.,  2002, \mn@doi [\mnras] {10.1046/j.1365-8711.2002.05990.x}, \href
  {https://ui.adsabs.harvard.edu/abs/2002MNRAS.336L..33R} {336, L33}

\bibitem[\protect\citeauthoryear{{Rivera-Thorsen} et~al.,}{{Rivera-Thorsen}
  et~al.}{2017}]{RiveraThorsen2017bSunburst}
{Rivera-Thorsen} T.~E.,  et~al., 2017, \mn@doi [\aap]
  {10.1051/0004-6361/201732173}, \href
  {https://ui.adsabs.harvard.edu/abs/2017A&A...608L...4R} {608, L4}

\bibitem[\protect\citeauthoryear{{Rivera-Thorsen} et~al.,}{{Rivera-Thorsen}
  et~al.}{2019}]{rivera2019gravitational}
{Rivera-Thorsen} T.~E.,  et~al., 2019, \mn@doi [Science]
  {10.1126/science.aaw0978}, \href
  {https://ui.adsabs.harvard.edu/abs/2019Sci...366..738R} {366, 738}

\bibitem[\protect\citeauthoryear{{Rodney} et~al.,}{{Rodney}
  et~al.}{2018}]{2018NatAs...2..324R}
{Rodney} S.~A.,  et~al., 2018, \mn@doi [Nature Astronomy]
  {10.1038/s41550-018-0405-4}, \href
  {https://ui.adsabs.harvard.edu/\#abs/2018NatAs...2..324R} {2, 324}

\bibitem[\protect\citeauthoryear{{Rosdahl} et~al.,}{{Rosdahl}
  et~al.}{2018}]{Rosdahl2018binary}
{Rosdahl} J.,  et~al., 2018, \mn@doi [\mnras] {10.1093/mnras/sty1655}, \href
  {https://ui.adsabs.harvard.edu/abs/2018MNRAS.479..994R} {479, 994}

\bibitem[\protect\citeauthoryear{{Sana}}{{Sana}}{2017}]{sana2016multiplicity}
{Sana} H.,  2017, in {Eldridge} J.~J.,  {Bray} J.~C.,  {McClelland} L.~A.~S.,
  {Xiao} L.,  eds,  IAU Symposium Vol. 329, The Lives and Death-Throes of
  Massive Stars. pp 110--117 (\mn@eprint {arXiv} {1703.01608}),
  \mn@doi{10.1017/S1743921317003209}

\bibitem[\protect\citeauthoryear{{Sana} \& {Evans}}{{Sana} \&
  {Evans}}{2011}]{sana2010multiplicity}
{Sana} H.,  {Evans} C.~J.,  2011, in {Neiner} C.,  {Wade} G.,  {Meynet} G.,
  {Peters} G.,  eds,  IAU Symposium Vol. 272, Active OB Stars: Structure,
  Evolution, Mass Loss, and Critical Limits. pp 474--485 (\mn@eprint {arXiv}
  {1009.4197}), \mn@doi{10.1017/S1743921311011124}

\bibitem[\protect\citeauthoryear{{Sana} et~al.,}{{Sana}
  et~al.}{2012}]{Sana2012MassiveBinaries}
{Sana} H.,  et~al., 2012, \mn@doi [Science] {10.1126/science.1223344}, \href
  {https://ui.adsabs.harvard.edu/abs/2012Sci...337..444S} {337, 444}

\bibitem[\protect\citeauthoryear{{Sana} et~al.,}{{Sana}
  et~al.}{2013}]{Sana2013MultipleOStars}
{Sana} H.,  et~al., 2013, \mn@doi [\aap] {10.1051/0004-6361/201219621}, \href
  {https://ui.adsabs.harvard.edu/abs/2013A&A...550A.107S} {550, A107}

\bibitem[\protect\citeauthoryear{{Sasaki}, {Suyama}, {Tanaka}  \&
  {Yokoyama}}{{Sasaki} et~al.}{2016}]{sasaki2016primordial}
{Sasaki} M.,  {Suyama} T.,  {Tanaka} T.,   {Yokoyama} S.,  2016, \mn@doi [\prl]
  {10.1103/PhysRevLett.117.061101}, \href
  {https://ui.adsabs.harvard.edu/abs/2016PhRvL.117f1101S} {117, 061101}

\bibitem[\protect\citeauthoryear{{Schechter} \& {Wambsganss}}{{Schechter} \&
  {Wambsganss}}{2002}]{schechter2002quasar}
{Schechter} P.~L.,  {Wambsganss} J.,  2002, \mn@doi [\apj] {10.1086/343856},
  \href {https://ui.adsabs.harvard.edu/abs/2002ApJ...580..685S} {580, 685}

\bibitem[\protect\citeauthoryear{{Schlafly} \& {Finkbeiner}}{{Schlafly} \&
  {Finkbeiner}}{2011}]{schlafly2011measuring}
{Schlafly} E.~F.,  {Finkbeiner} D.~P.,  2011, \mn@doi [\apj]
  {10.1088/0004-637X/737/2/103}, \href
  {https://ui.adsabs.harvard.edu/abs/2011ApJ...737..103S} {737, 103}

\bibitem[\protect\citeauthoryear{{Schneider}}{{Schneider}}{1987}]{1987ApJ...319....9S}
{Schneider} P.,  1987, \mn@doi [\apj] {10.1086/165428}, \href
  {https://ui.adsabs.harvard.edu/abs/1987ApJ...319....9S} {319, 9}

\bibitem[\protect\citeauthoryear{{Schneider} \& {Weiss}}{{Schneider} \&
  {Weiss}}{1988}]{SchneiderWeiss1988LightProp}
{Schneider} P.,  {Weiss} A.,  1988, \mn@doi [\apj] {10.1086/166214}, \href
  {https://ui.adsabs.harvard.edu/abs/1988ApJ...327..526S} {327, 526}

\bibitem[\protect\citeauthoryear{{Seitz} \& {Schneider}}{{Seitz} \&
  {Schneider}}{1994}]{Seitz1994microlensingI}
{Seitz} C.,  {Schneider} P.,  1994, \aap, \href
  {https://ui.adsabs.harvard.edu/abs/1994A&A...288....1S} {288, 1}

\bibitem[\protect\citeauthoryear{{Seitz}, {Wambsganss}  \& {Schneider}}{{Seitz}
  et~al.}{1994}]{Seitz1994microlensingII}
{Seitz} C.,  {Wambsganss} J.,   {Schneider} P.,  1994, \aap, \href
  {https://ui.adsabs.harvard.edu/abs/1994A&A...288...19S} {288, 19}

\bibitem[\protect\citeauthoryear{{Smith}}{{Smith}}{2017}]{Smith2017LBVreview}
{Smith} N.,  2017, \mn@doi [Philosophical Transactions of the Royal Society of
  London Series A] {10.1098/rsta.2016.0268}, \href
  {https://ui.adsabs.harvard.edu/abs/2017RSPTA.37560268S} {375, 20160268}

\bibitem[\protect\citeauthoryear{{Stanway} \& {Eldridge}}{{Stanway} \&
  {Eldridge}}{2018}]{stanway2018re}
{Stanway} E.~R.,  {Eldridge} J.~J.,  2018, \mn@doi [\mnras]
  {10.1093/mnras/sty1353}, \href
  {https://ui.adsabs.harvard.edu/abs/2018MNRAS.479...75S} {479, 75}

\bibitem[\protect\citeauthoryear{{Stanway}, {Eldridge}  \& {Becker}}{{Stanway}
  et~al.}{2016}]{Stanway2016BinaryReionization}
{Stanway} E.~R.,  {Eldridge} J.~J.,   {Becker} G.~D.,  2016, \mn@doi [\mnras]
  {10.1093/mnras/stv2661}, \href
  {https://ui.adsabs.harvard.edu/abs/2016MNRAS.456..485S} {456, 485}

\bibitem[\protect\citeauthoryear{{Sunyaev} \& {Zeldovich}}{{Sunyaev} \&
  {Zeldovich}}{1972}]{1972CoASP...4..173S}
{Sunyaev} R.~A.,  {Zeldovich} Y.~B.,  1972, Comments on Astrophysics and Space
  Physics, \href {https://ui.adsabs.harvard.edu/abs/1972CoASP...4..173S} {4,
  173}

\bibitem[\protect\citeauthoryear{{Tisserand} et~al.,}{{Tisserand}
  et~al.}{2007}]{tisserand2007limits}
{Tisserand} P.,  et~al., 2007, \mn@doi [\aap] {10.1051/0004-6361:20066017},
  \href {https://ui.adsabs.harvard.edu/abs/2007A&A...469..387T} {469, 387}

\bibitem[\protect\citeauthoryear{{Tuntsov}, {Lewis}, {Ibata}  \&
  {Kneib}}{{Tuntsov} et~al.}{2004}]{Tuntsov2004ClusterMicrolensing}
{Tuntsov} A.~V.,  {Lewis} G.~F.,  {Ibata} R.~A.,   {Kneib} J.~P.,  2004,
  \mn@doi [\mnras] {10.1111/j.1365-2966.2004.08115.x}, \href
  {https://ui.adsabs.harvard.edu/abs/2004MNRAS.353..853T} {353, 853}

\bibitem[\protect\citeauthoryear{{Umetsu}}{{Umetsu}}{2020}]{Umetsu2020ClusterLensingReview}
{Umetsu} K.,  2020, arXiv e-prints, \href
  {https://ui.adsabs.harvard.edu/abs/2020arXiv200700506U} {p. arXiv:2007.00506}

\bibitem[\protect\citeauthoryear{{Vanzella} et~al.,}{{Vanzella}
  et~al.}{2017}]{Vanzella2017SuperStarCluster}
{Vanzella} E.,  et~al., 2017, \mn@doi [\apj] {10.3847/1538-4357/aa74ae}, \href
  {https://ui.adsabs.harvard.edu/abs/2017ApJ...842...47V} {842, 47}

\bibitem[\protect\citeauthoryear{{Vanzella} et~al.,}{{Vanzella}
  et~al.}{2020a}]{vanzella2020transient}
{Vanzella} E.,  et~al., 2020a, \mn@doi [\mnras] {10.1093/mnrasl/slaa163}, \href
  {https://ui.adsabs.harvard.edu/abs/2020MNRAS.tmpL.182V} {}

\bibitem[\protect\citeauthoryear{{Vanzella} et~al.,}{{Vanzella}
  et~al.}{2020b}]{vanzella2020ionizing}
{Vanzella} E.,  et~al., 2020b, \mn@doi [\mnras] {10.1093/mnras/stz2286}, \href
  {https://ui.adsabs.harvard.edu/abs/2020MNRAS.491.1093V} {491, 1093}

\bibitem[\protect\citeauthoryear{{Venumadhav}, {Dai}  \&
  {Miralda-Escud{\'e}}}{{Venumadhav} et~al.}{2017}]{2017ApJ...850...49V}
{Venumadhav} T.,  {Dai} L.,   {Miralda-Escud{\'e}} J.,  2017, \mn@doi [\apj]
  {10.3847/1538-4357/aa9575}, \href
  {http://adsabs.harvard.edu/abs/2017ApJ...850...49V} {850, 49}

\bibitem[\protect\citeauthoryear{{Vietri} \& {Ostriker}}{{Vietri} \&
  {Ostriker}}{1983}]{1983ApJ...267..488V}
{Vietri} M.,  {Ostriker} J.~P.,  1983, \mn@doi [\apj] {10.1086/160886}, \href
  {https://ui.adsabs.harvard.edu/abs/1983ApJ...267..488V} {267, 488}

\bibitem[\protect\citeauthoryear{{Wambsganss}}{{Wambsganss}}{1999}]{wambsganss1999gravitational}
{Wambsganss} J.,  1999, Journal of Computational and Applied Mathematics, \href
  {https://ui.adsabs.harvard.edu/abs/1999JCoAM.109..353W} {109, 353}

\bibitem[\protect\citeauthoryear{{Wambsganss}}{{Wambsganss}}{2001}]{Wambsganss2000CosmoMicrolensing}
{Wambsganss} J.,  2001, in {Menzies} J.~W.,  {Sackett} P.~D.,  eds,
  Astronomical Society of the Pacific Conference Series Vol. 239, Microlensing
  2000: A New Era of Microlensing Astrophysics. p.~351 (\mn@eprint {arXiv}
  {astro-ph/0008419})

\bibitem[\protect\citeauthoryear{{Witt}}{{Witt}}{1993}]{Witt1993EfficientMethod}
{Witt} H.~J.,  1993, \mn@doi [\apj] {10.1086/172223}, \href
  {https://ui.adsabs.harvard.edu/abs/1993ApJ...403..530W} {403, 530}

\bibitem[\protect\citeauthoryear{{Wyithe} \& {Turner}}{{Wyithe} \&
  {Turner}}{2001}]{wyithe2001determining}
{Wyithe} J.~S.~B.,  {Turner} E.~L.,  2001, \mn@doi [\mnras]
  {10.1046/j.1365-8711.2001.03917.x}, \href
  {https://ui.adsabs.harvard.edu/abs/2001MNRAS.320...21W} {320, 21}

\bibitem[\protect\citeauthoryear{{Wyithe} \& {Turner}}{{Wyithe} \&
  {Turner}}{2002}]{WyitheTurner2002quasarGRBvariability}
{Wyithe} J.~S.~B.,  {Turner} E.~L.,  2002, \mn@doi [\apj] {10.1086/341481},
  \href {https://ui.adsabs.harvard.edu/abs/2002ApJ...575..650W} {575, 650}

\bibitem[\protect\citeauthoryear{{Young}}{{Young}}{1981}]{Young1981QSOMicrolensing}
{Young} P.,  1981, \mn@doi [\apj] {10.1086/158752}, \href
  {https://ui.adsabs.harvard.edu/abs/1981ApJ...244..756Y} {244, 756}

\bibitem[\protect\citeauthoryear{{Zibetti}, {White}, {Schneider}  \&
  {Brinkmann}}{{Zibetti} et~al.}{2005}]{zibetti2005intergalactic}
{Zibetti} S.,  {White} S. D.~M.,  {Schneider} D.~P.,   {Brinkmann} J.,  2005,
  \mn@doi [\mnras] {10.1111/j.1365-2966.2005.08817.x}, \href
  {https://ui.adsabs.harvard.edu/abs/2005MNRAS.358..949Z} {358, 949}

\bibitem[\protect\citeauthoryear{{Zumalac{\'a}rregui} \&
  {Seljak}}{{Zumalac{\'a}rregui} \& {Seljak}}{2018}]{zumalacarregui2018limits}
{Zumalac{\'a}rregui} M.,  {Seljak} U.,  2018, \mn@doi [\prl]
  {10.1103/PhysRevLett.121.141101}, \href
  {https://ui.adsabs.harvard.edu/abs/2018PhRvL.121n1101Z} {121, 141101}

\bibitem[\protect\citeauthoryear{{Zwicky}}{{Zwicky}}{1951}]{zwicky1951coma}
{Zwicky} F.,  1951, \mn@doi [\pasp] {10.1086/126318}, \href
  {https://ui.adsabs.harvard.edu/abs/1951PASP...63...61Z} {63, 61}

\makeatother
\end{thebibliography}

\label{lastpage}
\end{document}